\documentclass[twocolumn,journal]{IEEEtran}
\usepackage[latin9]{inputenc}
\usepackage{color}
\usepackage{array}
\usepackage{verbatim}
\usepackage{float}
\usepackage{amsmath}
\usepackage{amssymb}
\usepackage{graphicx}
\usepackage[numbers,numbers,sort&compress]{natbib}

\makeatletter

\providecommand{\tabularnewline}{\\}

\let\oldforeign@language\foreign@language
\DeclareRobustCommand{\foreign@language}[1]{%
  \lowercase{\oldforeign@language{#1}}}

\usepackage[ruled,vlined,linesnumbered]{algorithm2e}

\SetKwInput{KwInput}{Input}
\SetKwInput{KwOutput}{Output}

\usepackage{algorithmicx}
\usepackage{algpseudocode}
\usepackage{amsmath,amsfonts,amssymb}

\usepackage{array,booktabs,arydshln}
\usepackage{xcolor}
\usepackage{arydshln}

\usepackage{float} 

\usepackage[caption=false,font=footnotesize]{subfig}

\@ifundefined{showcaptionsetup}{}{%
 \PassOptionsToPackage{caption=false}{subfig}}
\usepackage{subfig}
\makeatother

\begin{document}
\title{A Coverage-Aware Resource Provisioning \\
Method for Network Slicing\thanks{Parts of this work have been presented at IEEE GLOBECOM 2018 and IEEE
ICC 2020.}}
\author{Quang-Trung~Luu,~\IEEEmembership{Student Member,~IEEE,} Sylvaine~Kerboeuf,
Alexandre~Mouradian, \\
and~Michel~Kieffer,~\IEEEmembership{Senior Member,~IEEE} \thanks{Q.-T.~Luu is with Nokia Bell Labs and the L2S, CNRS-CentraleSup\'elec-Univ
Paris-Sud-Univ Paris-Saclay, France, e-mail: quang\_ trung.luu@nokia.com.}\thanks{S.~Kerboeuf is with Nokia Bell Labs, France, e-mail: sylvaine.kerboeuf@
nokia-bell-labs.com.}\thanks{A.~Mouradian and M.~Kieffer are with the L2S, CNRS-CentraleSup\'elec-Univ
Paris-Sud-Univ Paris-Saclay, France, e-mail: \{michel.kieffer, alexandre.mouradian\}@l2s.centralesupelec.fr.}}
\markboth{Accepted to IEEE/ACM Transactions on Networking}{Q.-T.~Luu \MakeLowercase{\emph{et al.}}: Your Title}
\maketitle
\begin{abstract}
With network slicing in 5G networks, Mobile Network Operators can
create various slices for Service Providers (SPs) to accommodate customized
services. Usually, the various Service Function Chains (SFCs) belonging
to a slice are deployed on a best-effort basis. Nothing ensures that
the Infrastructure Provider (InP) will be able to allocate enough
resources to cope with the increasing demands of some SP. Moreover,
in many situations, slices have to be deployed over some geographical
area: coverage as well as minimum per-user rate constraints have then
to be taken into account.

This paper takes the InP perspective and proposes a slice resource
\emph{provisioning} approach to cope with multiple slice demands in
terms of computing, storage, coverage, and rate constraints. The resource
requirements of the various SFCs within a slice are aggregated within
a graph of Slice Resource Demands (SRD). Infrastructure nodes and
links have then to be provisioned so as to satisfy all SRDs. This
problem leads to a Mixed Integer Linear Programming formulation. A
two-step approach is considered, with several variants, depending
on whether the constraints of each slice to be provisioned are taken
into account sequentially or jointly. Once provisioning has been performed,
any slice deployment strategy may be considered on the reduced-size
infrastructure graph on which resources have been provisioned. Simulation
results demonstrate the effectiveness of the proposed approach compared
to a more classical direct slice embedding approach.
\end{abstract}


\IEEEpeerreviewmaketitle{}

\section{Introduction}

\IEEEPARstart{N}{etwork} Function Virtualization (NFV) is attracting
widespread interest due to the overall equipment and management cost
reductions it allows \cite{Liang2014} and to the increased network
flexibility it provides \cite{Basta2014}. Using NFV, network functions
are decoupled from their hosting hardware and are offered as virtualized
services decomposed in \textit{Virtual Network Functions} (VNFs) on
general-purpose servers. With cloud networks, infrastructure is also
evolving to integrate edge and central data centers onto which VNFs
may be deployed using IT technologies. With the help of virtualization,
many dedicated end-to-end network services can co-exist and share
the same physical infrastructure, while relying on different network
capabilities, protocols, and network architecture optimized towards
customized requirements. The network slicing concept has thus emerged
in 5G networks \cite{5GAmericas2016,IETF2017,Rost2017}. Slicing can
be applied for deploying business cases such as multi-tenants sharing
the same network infrastructure, where tenants, \textit{i.e.}, vertical
actors, can operate and manage their own network slice to address
applications in energy, e-health, smart city, connected cars \cite{IETF2017}.

A network slice can be seen as a collection of \textit{Service Function
Chains} (SFCs) and a set of physical network resources, which are
dynamically allocated to build a customized logically isolated virtual
network. Each SFC consists of several interconnected VNFs describing
the processing applied to a data flow related to a given service.
With cloudification technology, SFCs and VNFs can be easily and flexibly
initialized, launched, chained, and scaled to meet changeable workload
requests \cite{Luong2018}. Iterative SFC deployment strategies are
well-suited to such dynamic slice management. Nevertheless, when several
concurrent slices are managed in parallel, nothing ensures that enough
infrastructure resources will be available to deploy a new SFC. Such
best-effort slice management makes it difficult to satisfy a \textit{Service
Level Agreement} (SLA) expressed by tenants in terms, \textit{e.g.},
of guaranteed amount of serviced users. More flexible models and mechanisms
for network service provisioning and deployment are needed \cite{DeSousa2018}.
Additionally, research challenges remain when network slicing incorporates
the wireless part of legacy or 5G networks \cite{Li2017,Kaloxylos2018},
where multiple network segments including the radio access, transport,
and core network, have to be considered.

This paper studies the way to efficiently provision and deploy end-to-end
network slices on radio and cloud network infrastructures in a multi-tenancy
context. As in \cite{Luu2018}, our work focuses on the problem of
slice resource provisioning, \emph{i.e.}, reservation. By provisioning
we ensure that enough resource is reserved for further SFC deployment
while satisfying coverage constraints for mobile end-users of the
slice services. A two-step method is proposed for efficient slice
deployment: The resource provisioning process is followed by the SFC
embedding process. For the latter any state-of-the-art deployment
approach may be employed on a simplified infrastructure network reduced
to the nodes and links which have provisioned resources. The SFC embedding
time may then be much smaller. We extend preliminary results obtained
in \cite{Luu2018}, by accounting for radio coverage constraints.
This requires the introduction of a radio propagation model in the
radio resource provisioning phase. Moreover, coverage requirements
impose several additional constraints on the SFCs to be deployed within
the network infrastructure.

The rest of the paper is structured as follows. Section~\ref{sec:Related-Work}
presents the system architecture, analyzes related work, and highlights
our main contributions. The model of the infrastructure network and
of the slice resource demands are presented in Section~\ref{sec:NetworkModel}.
The slice resource provisioning problem is then formulated in Section~\ref{sec:ProblemFormulation}
as a mixed integer linear programming problem accounting for cloud
network and radio resource constraints for the deployment of multiple
slices. An optimal and four suboptimal variants of a coverage-aware
slice resource provisioning algorithm are provided in Section~\ref{sec:Single-step-vs-Two-step}.
Numerical results are presented in Section~\ref{sec:Evaluation}.
Finally, Section~\ref{sec:Conclusions} draws some conclusions and
perspectives.

\section{System Architecture, Related Work, and Main Contributions\label{sec:Related-Work}}

\subsection{System Architecture}

\begin{figure}[tbh]
\begin{centering}
\includegraphics[width=1\columnwidth]{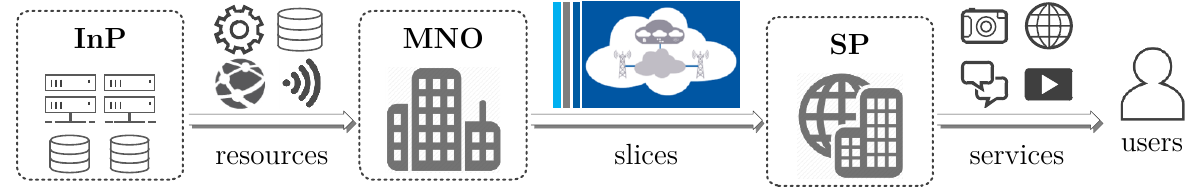}
\par\end{centering}
\caption{System architecture \label{fig:System-architecture}}
\end{figure}

Several entities are involved in network slicing, as described in
Figure~\ref{fig:System-architecture} \cite{Liang2014}. The \textit{Infrastructure
Providers} (InPs) own and manage the wireless and wired infrastructure
such as the cell sites, the fronthaul and backhaul networks, and cloud
data centers.

The \textit{Mobile Network Operator} (MNO) leases resources from InPs
to setup and manage the slices. The \textit{Service Providers} (SPs)
exploit the slices supplied by the MNO, and provide to their customers
the required services that are running within the slices. Service
needs are forwarded by the SP to the MNO within an SLA. The SLA describes,
at a high level of abstraction, characteristics of the service with
the desired QoS, the number of devices (or the device/user density),
the geographical region where the service has to be made available
for the end-users, \textit{etc}. Due to user mobility, these characteristics
may be time-varying. The MNO translates the SP high-level demands
into SFCs able to fulfill the service requirements. SFCs are then
deployed on the network infrastructure so that QoS requirements are
satisfied.

In this paper, one considers an infrastructure owned by a single InP.
To perform this deployment, the InP has to identify the infrastructure
nodes on which the VNFs are deployed and the links able to transmit
data between these nodes. Given a set of SFC demands, this consists
in finding \textit{i}) Base Stations (BS) providing radio resources
to mobile users so as to satisfy coverage constraints, \emph{ii})
the placement of the VNFs on the data center nodes, and \textit{iii})
the routing of data flows between the VNFs, while respecting the structure
of SFCs and optimizing a given objective (\textit{e.g}., minimizing
the infrastructure and software fees cost). Updates may be necessary
when the service characteristics have changed significantly.

Our aim, with resource provisioning is to reserve, somewhat in advance,
enough infrastructure resources to ensure that the MNO will have access
to properly located radio resources and be able to deploy the set
of SFCs with characteristics as stated in the SLA. The time scale
at which provisioning is performed is much larger than that at which
SFCs are deployed and adapted to meet actual time-varying user demands.
One focuses on a time interval over which resources will be provisioned
so as to be compliant with the variations of user demands within a
slice. The duration of this time interval results from a compromise
between the need to update the provisioning and the level of conservatism
in the amount of provisioned resources required to satisfy fast fluctuating
user demands.

\begin{figure}[tbh]
\begin{centering}
\includegraphics[width=0.8\columnwidth]{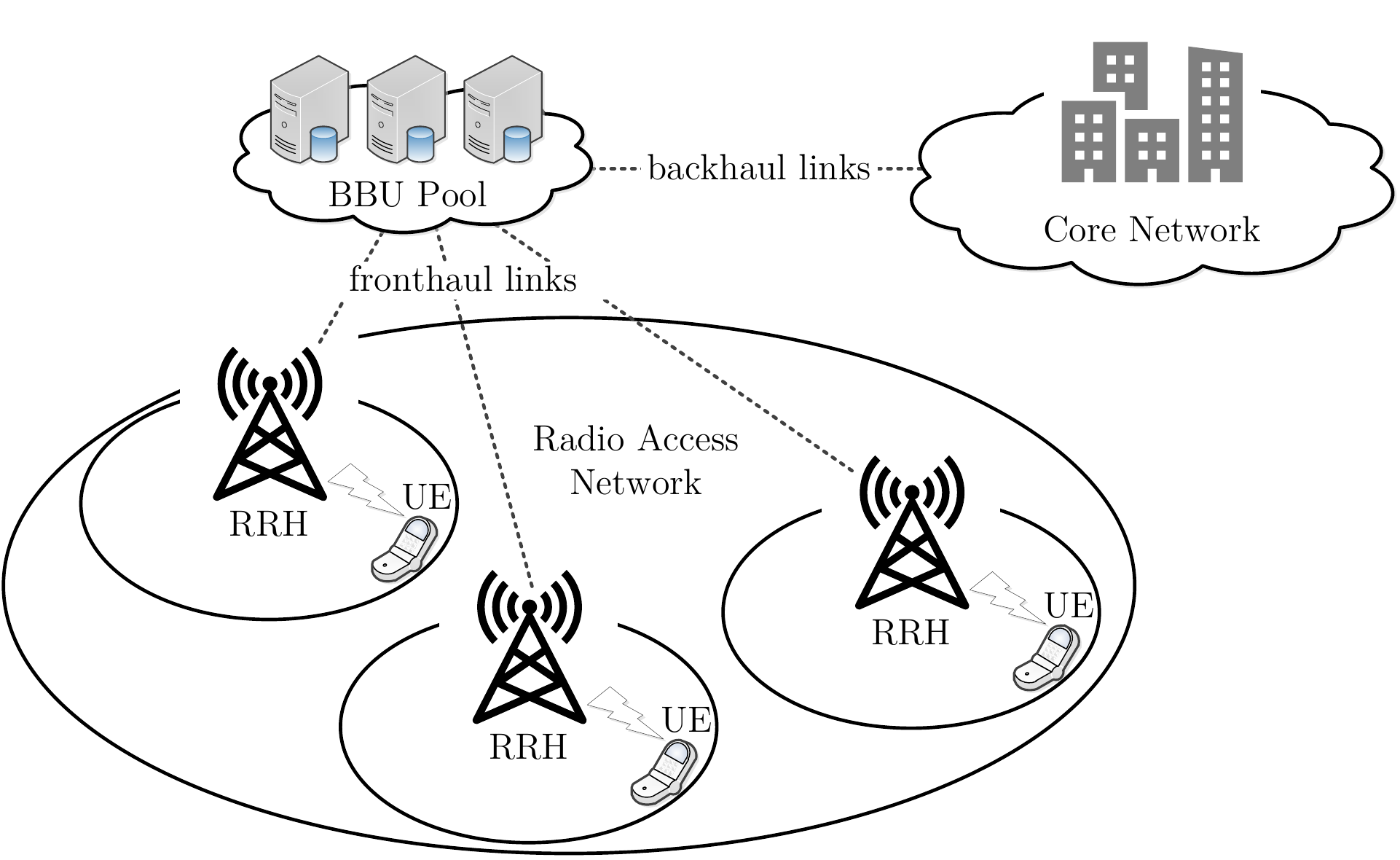}
\par\end{centering}
\caption{General architecture of C-RAN \label{fig:CRAN-architecture}}
\end{figure}

\textit{\emph{In this work, we adopt the }}\emph{Cloud Radio Access
Network}\textit{\emph{ (C-RAN) architecture, a cloud architecture
for future mobile network, illustrated in Figure~\ref{fig:CRAN-architecture}.
The C-RAN nodes (}}\emph{i.e}\textit{\emph{., eNB for 4G and gNB for
5G) mainly consists of two parts: The distributed }}\emph{Remote Radio
Heads}\textit{\emph{ (RRHs) plus antennas deployed at the cellular
radio sites and the centralized }}\emph{Base Band Unit}\textit{\emph{
(BBU) pool hosted in an edge cloud data center \cite{Tran2017}. The
BBU pool hosts multiple virtual BBUs and handles higher layer processing
functions, whereas all basic radio functions remain at the cellular
radio station with the RRH. In 4G, the BBU handles all the L1-L2-L3
functional layers whereas radio frequency functions reside at the
RRH. Within 5G, the gNB is split in three parts, namely }}\emph{Central
Unit}\textit{\emph{ (CU), }}\emph{Distributed Unit}\textit{\emph{
(DU), and }}\emph{Radio Unit}\textit{\emph{ (RU), and different functional
splits are under study where in some options the RU can support some
L2 functions thus reducing the capacity required for the fronthaul
link \cite{ITU-T2018}. The link (interface) between the BBU and the
RRH is known as the fronthaul whereas the backhaul network connects
the BBU with the core network functions hosted in the regional or
central cloud.}}

\subsection{Related Work}

Early results on assigning infrastructure network resources to virtual
network components may be found, \emph{e.g.}, in \cite{Zhu2006,Chowdhury2012}.
Due to its capability of sharing efficiently network resources in
5G networks, the concept of network virtualization has gained renewed
attention in the literature \cite{Rost2017,Foukas2017,Nakao2017,Afolabi2018}
via the concept of network slicing.

Network slice resource allocation is a complex problem. When a slice
instance is seen as a collection of SFCs, slice embedding needs to
deploy the SFCs on a shared infrastructure while satisfying various
constraints. Most of prior works related to SFC and VNF deployment
do not account for coverage constraints. For example, in \cite{Riggio2016,Vizarreta2017},
computing, storage, and aggregated wireless resource demands of SFCs
are considered. The minimization of the SFC embedding cost is formulated
either as an \textit{Integer Linear Programming} (ILP) \cite{Vizarreta2017,Cohen2015,Riera2016}
or as a \textit{Mixed Integer Linear Programming} (MILP) problem \cite{Chowdhury2012,Kang2017},
which are known to be NP-hard \cite{Fischer2013}. In \cite{Tajiki2018},
the VNF placement problem is expressed as an \textit{Integer Quadratic
Programming} (IQP) problem with a set of energy consumption constraints,
and then is transformed to a solvable linear form.

To address the high computational complexity resulting from the ILPs
or MILPs, various heuristics have been proposed, see, \emph{e.g.},
\cite{Riggio2016,Vizarreta2017,Cohen2015}. For example, \cite{Riggio2016}
introduced a heuristic based on the search of shortest paths to sequentially
embed the SFCs. In \cite{Vizarreta2017}, the candidate infrastructure
nodes are sorted to find the best node, in terms of deployment cost,
to host a given VNF. Its neighbors are then considered as candidates
to deploy the next VNF.

The \textit{Column Generation} (CG) technique has been widely studied
to solve large ILP problems \cite{Huin2017}. With CG, the original
ILP is decomposed into a \textit{Master Problem} (MP) and a \textit{Pricing
Problem} (PP). The MP is the original problem where only a subset
of variables is considered. The PP is a new problem created to identify
a new variable\textit{, i.e}., a column, to add to the MP to improve
the current solution. In \cite{Huin2017} or \cite{Liu2017}, CG has
been used to relax ILP-based SFC embedding or reconfiguration problems.
Specifically, in \cite{Huin2017}, the SFC embedding problem is addressed.
Only core capacity and bandwidth resources for infrastructure nodes
and links are considered. In \cite{Liu2017}, the embedding of new
SFCs and the re-adjustment of in-service SFCs are both considered.
Re-adjustment of in-service SFCs may imply the migration of VNFs and
virtual links may need to be updated to meet changes of resource demands.
This problem is again formulated as an ILP where the objective is
to minimize the deployment as well as the migration costs. Only linear
SFCs are allowed and any node with radio resource may serve as access
point for the users, which makes difficult the satisfaction of coverage
constraints. Moreover, possible paths in the network are assumed to
be available, which needs some computational effort before the deployment.

In \cite{Mechtri2016}, the join VNF and virtual link placement is
formulated as a \textit{Weighted Graph Matching Problem} (WGMP), where
the SFC graph and the infrastructure graph are modeled as weighted
graphs, on which each node and each link have their own weight corresponding
to their required resource (for the SFC graph), or their available
resource (for the infrastructure graph). An \textit{\emph{eigendecomposition}}-based
method is then proposed to solve the WGMP problem, whose aim is to
find, with a reduced complexity, the optimum matching between the
SFC graph and the infrastructure graph. In \cite{Huin2017}, \cite{Liu2017},
and \cite{Mechtri2016}, a unique type of resource is considered at
infrastructure nodes (processing) and at links (bandwidth). Radio
resource is not considered.

The resource allocation problem among competing slices in a heterogeneous
cloud infrastructure is addressed in \cite{Halabian2019}. Slice resource
demands are aggregated in a vector of VNF resource demands in the
slice multiplied by a coefficient linked to the number of services
to be processed per time unit. The considered types of resource are
CPU, memory, bandwidth, and storage. The resource allocation among
multiple slices is performed considering two different approaches.
The first approach involves a centralized convex optimization problem,
whose objective is to maximize the total slice utility. Nevertheless,
as pointed out in \cite{Halabian2019}, such centralized solution
lacks of scalability, is not robust to a failure of the central optimizer,
and is prone to non-collaborative slice providers which may harm the
system. For these reasons, a distributed method based on game theory
is considered to improve robustness and scalability. Optimization
is performed in a decentralized way among the data centers and slice
providers. The results provided by all entities determines the final
resource allocation for all slices. Nevertheless, the placement of
VNFs in data centers is predetermined by the MNO and again, wireless
resources are not considered. A resource aggregation scheme similar
to that in \cite{Halabian2019} has been introduced in our previous
work \cite{Luu2018}, where infrastructure resources are provisioned
to satisfy slice resource demand constraints. Radio resources are
considered, but radio coverage constraints are still ignored.

The design of efficient allocation mechanisms for virtualized radio
resources has been recently addressed in \cite{Chatterjee2018}. This
paper aims at minimizing the leasing cost of BSs so as to meet SP
demands, while providing, with a given probability, a minimum data
rate for any user located in their coverage area. The rate constraint
is expressed as a linear function of the BS load (number of users
served by the BS), of the distance from users to the nearest BS, and
of the downlink interference. This linear approximation, however,
requires some assumptions. For instance, a user of an SP is assumed
to be served by its nearest BS among the set of BSs allocated to the
SP. This reduces somehow the potentiality of achieving the optimal
sharing of the radio resource.

In \cite{Teague}, a heterogeneous spatial user density is considered,
and the joint BS selection and adaptive slicing are formulated as
a two-stage stochastic optimization problem. The first stage aims
at defining the set of BSs to activate. The second stage aims at allocating
wireless resources of the BSs to each point of the region to be covered
by the SP. Several random realizations of user locations are generated
to get a reduced-complexity deterministic optimization problem. A
genetic algorithm is then used for the optimization.

In \cite{Lee2016}, a network slicing framework for \textit{\emph{multi-tenant
heterogeneous C-RAN}} is introduced. The sharing of radio resources
in terms of data rate is considered, with some constraints related
to the fronthaul capacity, the transmission power budget of RRHs,
or the tolerable interference threshold of an RRH on a sub-channel.
Slicing is formulated as a weighted throughput maximization problem,
which aims at maximizing the total rate obtained by users connected
to given RRHs on given sub-channels. Nevertheless, the proposed framework
does not consider computing and memory resources associated to the
processing within the BBUs. Such resources are assumed to be properly
scaled so as to support the required service rate. Moreover, the proposed
framework addresses only downlink data services.

The wireless network slicing problem is also addressed in \cite{DOro2018}.
A game theory-based distributed algorithm to solve the problem is
proposed. The proposed algorithm accounts for the limited availability
of wireless resources and considers different aspects such as congestion,
deployment costs and the RRH-user distance. The coverage area of RRH
is considered, but the possible coverage constraints required by the
slices are not taken into account.

\subsection{Main Contributions}

Compared to previous works, this paper considers slice resource demands
in terms of coverage and traffic requirements in the radio access
part of the network as well as network, storage, and computing requirements
from a cloud infrastructure of interconnected data centers for the
rest of the network. This work borrows the slice resource provisioning
approach introduced in \cite{Luu2018}, and adapts it to the joint
radio and network infrastructure resource provisioning. Constraints
related to the infrastructure network considered in \cite{Riggio2016,Vizarreta2017,Luu2018,Halabian2019}
are combined with coverage and radio resource constraints introduced
in \cite{Chatterjee2018,Teague,DOro2018,Lee2016}. The coverage constraints
are very important to satisfy mobile service requirements. The amount
of radio resources required depends on the location of users. A radio
propagation model is thus introduced in the provisioning phase. The
coverage constraints reduce the flexibility to select the nodes on
which SFCs are deployed.

In this work, we assume that the resource requirements for the various
SFCs that will have to be deployed within a slice may be aggregated
and represented by a Slice Resource Demand (SRD) graph that mimics
the graph of SFCs. These SRDs are evaluated by the MNO to satisfy
the QoS requirements imposed by the SP. The InP has then to provision
enough infrastructure resources to meet the SLA. Due to the fact that
nodes or links of the graph of SRDs represent aggregate requirements,
several infrastructure nodes may have to be gathered and parallel
physical links have to be considered to satisfy the various SRDs.
This is the main difference with respect to the traditional service
chain embedding approach considered for example in \cite{Riggio2016,Vizarreta2017},
where each VNF is deployed on a single node. In \cite{Riggio2016,Vizarreta2017},
virtual nodes and links are mapped on the infrastructure network to
allocate resources to VNFs and virtual links. In this paper, one provisions
a sufficient number of infrastructure nodes and links, so that the
aggregated provisioned resources meet the slice demands represented
by the graph of SRDs.

When provisioning slices, we consider coverage constraints, in which
slices are assumed to cover a specific region in the considered geographical
area, that is part of the SLA with the tenant. We devise the special
case of the cloud RAN architecture with RRHs which are nodes having
radio resources. In our model, radio resource blocks are allocated
and the channel between the RRH nodes and users is taken into account.
Compared with \cite{Chatterjee2018}, the selected BS is not necessarily
the nearest one. Moreover, both downlink and uplink traffic are considered
for the service rate model.

\section{System Model\label{sec:NetworkModel}}

\begin{figure}[tbh]
\begin{centering}
\includegraphics[width=1\columnwidth]{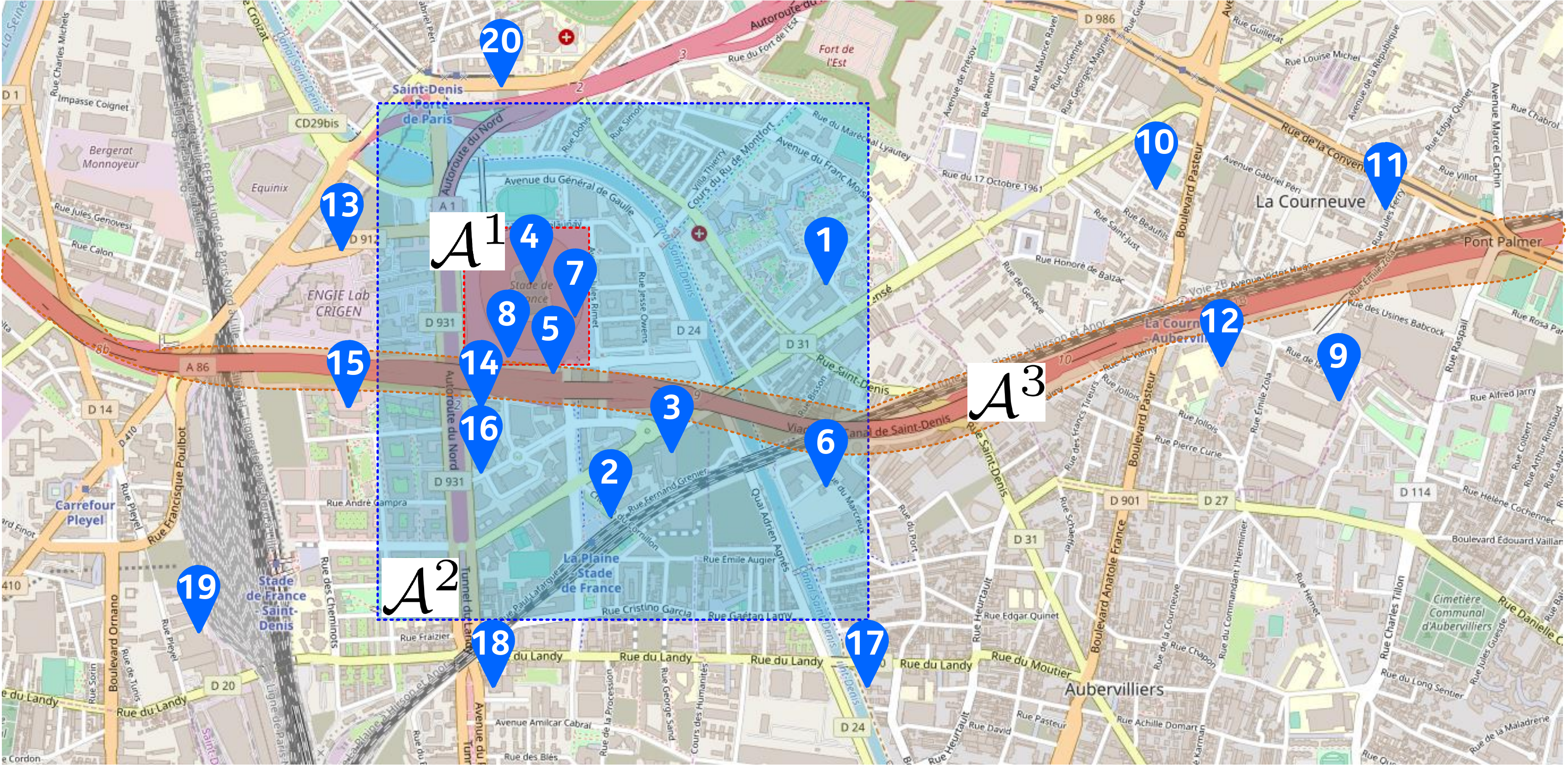}
\par\end{centering}
\caption{The considered metropolitan area including the Stade de France (covered
by the red rectangle representing $\mathcal{A}^{1}$), its surrounding
(blue rectangle representing $\mathcal{A}^{2}$), and part of the
A86 highway (orange shape representing $\mathcal{A}^{3}$). Blue markers
show the location of RRH nodes of Orange. \label{fig:Orange_Map}}
\end{figure}

Consider a set of SPs whose aim is to provide different services,
indexed by $\sigma=1,\dots,|\mathcal{S}|$, to mobile users. The geographical
area under study is denoted by $\mathcal{A}$ and the subarea over
which service $\sigma$ has to be made available is denoted by $\mathcal{\ensuremath{A}^{\sigma}}$.
For that purpose, each SP forwards his service requirements to the
MNO, whose aim is to design a network slice able to satisfy these
requirements. Figure~\ref{fig:Orange_Map} illustrates three typical
geographical subareas over which three different services have to
be deployed.

The MNO sends to the InP a Slice Resource Demand (SRD). This SRD consists
of (\textit{i}) an SRD graph accounting for the structure and SLA
of the slice, and (\textit{ii}) SRD coverage information related to
the area $\mathcal{A^{\sigma}}$ over which the service will have
to be made available. The InP is then in charge of provisioning enough
infrastructure resources to deploy the SFCs whose resource demands
have been described by the SRD graph.

This section details the model of the infrastructure provided by the
InP and the way a service with wireless coverage constraints can be
mapped to a slice with specific SRD graph.

\subsection{Infrastructure model}

Consider an infrastructure network managed by some InPs. This network
is represented by a directed graph $\mathcal{G}_{\textrm{I}}=\left(\mathcal{N}_{\textrm{I}},\mathcal{E}_{\textrm{I}}\right)$,
where $\mathcal{N}_{\textrm{I}}$ is the set of infrastructure nodes
and $\mathcal{E}_{\textrm{I}}$ is the set of infrastructure links,
which correspond to the wired connections between nodes and within
nodes (loopback links) of the infrastructure network.

Each infrastructure node $i\in\mathcal{N}_{\textrm{I}}$ is characterized
by a given amount of computing and storage resources, denoted as $a_{\text{c}}(i)$
and $a_{\text{s}}(i)$, which may be allocated to network slices.
Radio resources are exclusively provided by a subset $\mathcal{N}_{\textrm{Ir}}\subset\mathcal{N}_{\textrm{I}}$
of RRH nodes, whose location in some Cartesian frame attached to $\mathcal{A}$
is denoted by $x_{i}^{\textrm{r}}$. The cost associated to the use
of an infrastructure node~$i$ consists of a fixed part $c_{\text{f}}\left(i\right)$
for node disposal (paid by each slice using node~$i$) and a variable
part $c_{\text{c}}(i)$, $c_{\text{s}}(i)$, and $c_{\text{r}}(i)$,
which depend linearly on the amount of computing, storage, and radio
resources provided by that node.

Each infrastructure link $ij\in\mathcal{E}_{\textrm{I}}$ connecting
node $i$ to $j$ has a bandwidth $a_{\text{b}}(ij)$, and an associated
per-unit bandwidth cost $c_{\text{b}}(ij)$. Several distinct VNFs
of the same slice may be deployed on a given infrastructure node.
When communication between these VNFs is required, an internal (loopback)
infrastructure link $ii\in\mathcal{E}_{\textrm{I}}$ can be used at
each node $i\in\mathcal{N}_{\textrm{I}}$, as in \cite{Wang2009},
in the case of interconnected virtual machines (VMs) deployed on the
same host. The associated per-unit bandwidth cost, in that case, is
$c_{\textrm{\text{b}}}\left(ii\right)$.

\subsection{SRD Model}

An SRD is defined on the basis of an SLA between an SP and the MNO.
The SLA may consider several time intervals over each of which the
service characteristics and constraints are assumed constant, but
may vary from one interval to the next one. These time intervals translate,
\emph{e.g.}, day and night variations of user demands. They last between
tens of minutes and hours. It is of the responsibility of the SP and
MNO to properly scale the requirements expressed in the SLA, by considering,
for example, similar services deployed in the past.

In this paper, one considers a given time interval specified in the
SLA. The SLA is also expressed in terms of supported service type
and targeted QoS such as a minimum average data rate $\underline{R}_{\text{u}}^{\sigma}$
and $\underline{R}_{\text{d}}^{\sigma}$ for the wireless uplink and
downlink traffic of each client. The geographical distribution function
$\rho^{\sigma}(x)$, with $x\in\mathcal{A}$, describes the \emph{maximum}
user/device density to be served around $x$ within the considered
time interval. 

One assumes that the resource requirements for a slice can be represented
by an SRD graph that mimics the graph of SFCs. The SRD graph for slice
$\sigma$ is an oriented graph denoted by $\mathcal{G}_{\textrm{V}}^{\textrm{\ensuremath{\sigma}}}=(\mathcal{N}_{\textrm{V}}^{\textrm{\ensuremath{\sigma}}},\mathcal{E}_{\textrm{V }}^{\textrm{\ensuremath{\sigma}}})$,
where $\mathcal{N}_{\textrm{V}}^{\textrm{\ensuremath{\sigma}}}$ and
$\mathcal{E}_{\textrm{V }}^{\textrm{\ensuremath{\sigma}}}$ are respectively
the set of (virtual) SRD nodes and links. The SRD graph has a structure
close to the SFC graph, with SRD nodes corresponding to the VNFs of
the SFC. Each SRD node $v\in\mathcal{N}_{\textrm{V}}^{\textrm{\ensuremath{\sigma}}}$
is characterized by a given amount of \emph{required} computing and
storage resources, denoted as $r_{\textrm{c}}(v)$ and $r_{\textrm{s}}(v)$
to sustain the aggregated demand for all instances of a given VNF
in the slice. The minimum resources to deploy a single VNF instance
are denoted as $\underline{r}_{\textrm{c}}(v)$ and $\underline{r}_{\textrm{s}}(v)$.
Each link $vw\in\mathcal{E}_{\textrm{V }}^{\textrm{\ensuremath{\sigma}}}$,
connecting node $v$ to $w$ in the SRD graph, is characterized by
the bandwidth $r_{\text{b}}(vw)$ required to sustain the aggregated
traffic demand between the VNFs associated to $v$ and $w$.

SFCs will be deployed on the infrastructure nodes and links which
have provisioned resources. Enough resources should be provisioned
by each node to be able to host at least one VNF.

In the SRD graph, one assumes that the uplink and downlink radio resource
demands are associated to a single node $v_{\text{r}}$. The aggregated
uplink and downlink data rates $r_{\text{u}}^{\sigma}\left(v_{\text{r}}\right)$
and $r_{\text{d}}^{\sigma}\left(v_{\text{r}}\right)$ are associated
to the coverage constraint of slice $\sigma$
\begin{align}
 & r_{\text{u}}^{\sigma}\left(v_{\text{r}}\right)=\underline{R}_{\text{u}}^{\sigma}\int_{\mathcal{\mathcal{A}^{\sigma}}}\rho^{\sigma}\left(x\right)\mathrm{d}x,\nonumber \\
 & r_{\text{d}}^{\sigma}\left(v_{\text{r}}\right)=\underline{R}_{\text{d}}^{\sigma}\int_{\mathcal{\mathcal{A}^{\sigma}}}\rho^{\sigma}\left(x\right)\mathrm{d}x.\label{eq:SRDUpDownlink}
\end{align}

\begin{figure}[tbh]
\begin{centering}
\subfloat[Graph of SFCs.\label{fig:Branched_SFC}]{\begin{centering}
\includegraphics[width=1\columnwidth]{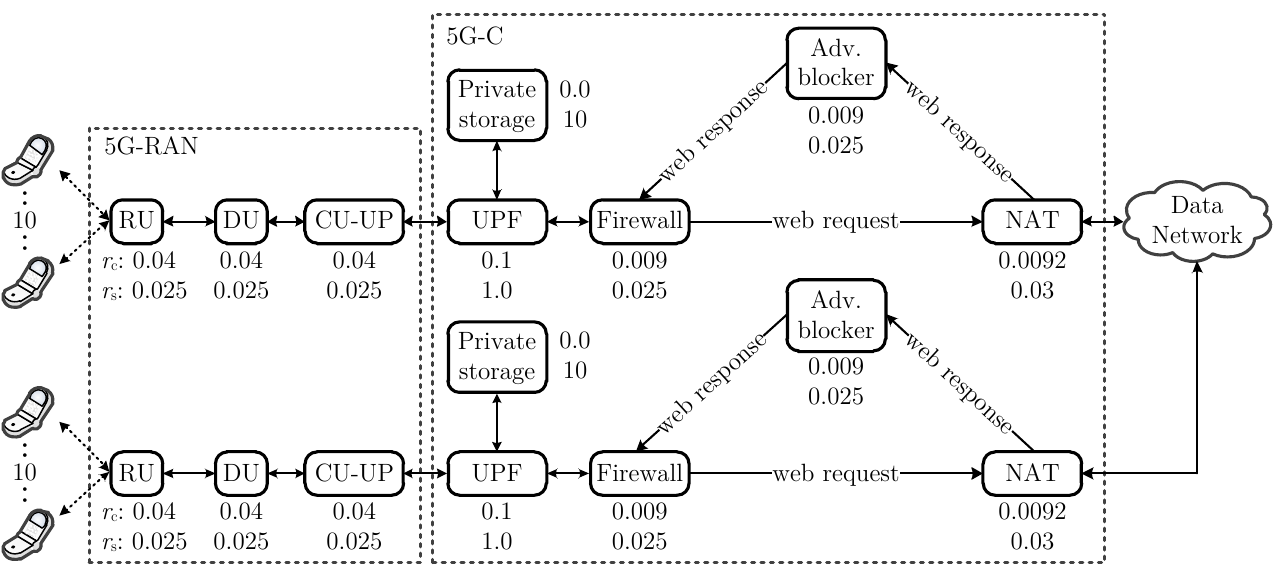}
\par\end{centering}
}
\par\end{centering}
\begin{centering}
\subfloat[Corresponding SRD graph.\label{fig:Branched_SRD}]{\begin{centering}
\includegraphics[width=1\columnwidth]{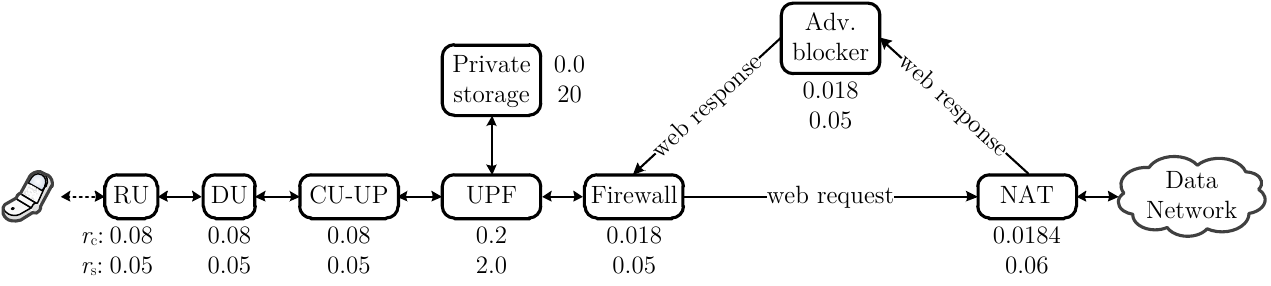}
\par\end{centering}
}
\par\end{centering}
\caption{SFCs and their required computing (in CPUs) and storage (in GBytes)
resources for the deployment of a secured web browsing service with
advertisement removal and their associated SRD graph.\label{fig:WebBrowsing}}
\end{figure}

Figure~\ref{fig:WebBrowsing} illustrates the SFCs required for the
deployment of a web browsing service with advertisement removal inspired
by \cite{Cerrato2018} and its associated SRD graph. Figure~\ref{fig:Branched_SFC}
describes the eight VNFs to be deployed, including: three RAN VNFs,
namely a RU to handle RF operations, a DU, and a \textit{Centralized
Unit for User-Plane} (CU-UP) to handle computing and processing loads;
and five VNFs placed in the core network, namely a \textit{User-Plane
Function} (UPF), a private storage management function, a firewall,
an advertisement blocker, and a \textit{Network Address Translation}
(NAT) function. Each of these VNFs is characterized by computing and
storage requirements. Some links are bidirectional, \emph{e.g.}, between
the UPF and the firewall, others are unidirectional, \emph{e.g.},
the uplink traffic from users has not to go through the advertisement
blocker. The corresponding SRD graph is represented in Figure~\ref{fig:Branched_SRD}.
All identical instances of SFCs deployed within the slice are represented
by a single graph whose structure is identical to the SFC graph. The
requirements in terms of storage, computing, and wireless capacity
of each component of the SRD graph aggregate the corresponding requirements
of the components of the SFC graph. More details are provided in Section~\ref{sec:Evaluation}.

\begin{figure}[tbh]
\begin{centering}
\subfloat[Graph of SFCs.\label{fig:Cyclic_SFC}]{\begin{centering}
\includegraphics[width=0.9\columnwidth]{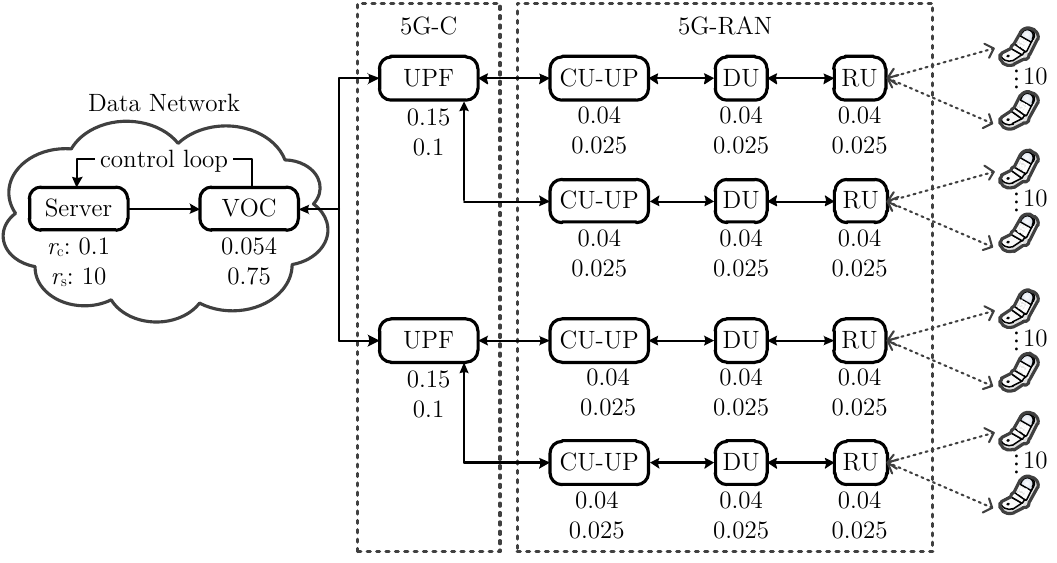}
\par\end{centering}
}
\par\end{centering}
\begin{centering}
\subfloat[Corresponding SRD graph.\label{fig:Cyclic_SRD}]{\begin{centering}
\includegraphics[width=0.9\columnwidth]{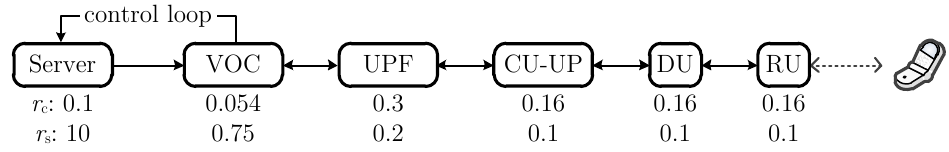}
\par\end{centering}
}
\par\end{centering}
\caption{SFCs and their required computing (in CPUs) and storage (in GBytes)
resources for the deployment of an adaptive wireless video streaming
service and their associated SRD graph. \label{fig:WirelessStreaming}}
\end{figure}

A second example is provided in Figure~\ref{fig:WirelessStreaming},
which represents the SFCs required for the deployment of an adaptive
wireless video streaming service and its associated SRD graph taken
from \cite{Savi2017}. Figure~\ref{fig:Cyclic_SFC} represents the
VNFs for the user-plane of the 5G-RAN (RU, DU, CU-UP), the 5G-Core
(UPF), and the server and\textit{ Video Optimization Controller} (VOC)
placed in the data network. The server archives videos with different
qualities (bitrate). Using the information received from users such
as the bandwidth or end-to-end latency, the VOC dynamically adjusts
the video bitrate to provide to the users. Figure~\ref{fig:Cyclic_SRD}
describes the associated SRD graph.

When it is possible to reserve enough resources, the MNO will be ensured
to be able to deploy a collection of SFCs needed to satisfy the SLA
over its time interval of validity. When, for example, the user density
over some subarea is larger than stated in the SLA, some users may
not be served. Nevertheless, from the perspective of the InP, the
SLA is still satisfied. On contrary, when the user density/requirements
are less than the maximum specified in the SLA, some provisioned resources
may remain unused, but this is the price to pay when provisioning
resources.

Table~\ref{tab:substrate_slice_parameter} summarizes all parameters
involved in the description of the infrastructure network and the
graph of SRDs for a slice.

\begin{table}[tbh]
\caption{Infrastructure Network and Slice Parameters. \label{tab:substrate_slice_parameter}}

\centering{}%
\begin{tabular}{>{\centering}p{0.12\columnwidth}>{\raggedright}p{0.7\columnwidth}}
\hline 
\multicolumn{2}{l}{\textbf{Node resource type: $n$}}\tabularnewline
$n$ & computing ($\textrm{c}$), storage ($\textrm{s}$), and radio ($\textrm{r}$)\tabularnewline
\hline 
\multicolumn{2}{l}{\textbf{Infrastructure network graph: $\mathcal{G}_{\textrm{I}}=\left(\mathcal{N}_{\textrm{I}},\mathcal{E}_{\textrm{I}}\right)$}}\tabularnewline
$\mathcal{N}_{\textrm{I}}$ & Set of infrastructure nodes\tabularnewline
$\mathcal{E}_{\textrm{I}}$ & Set of infrastructure links\tabularnewline
$a_{n}(i)$ & Available resource of type $n$ at node $i\in\mathcal{N}_{\textrm{I}}$\tabularnewline
$a_{\text{b}}(ij)$ & Available bandwidth of link $ij\in\mathcal{E}_{\textrm{I}}$\tabularnewline
$c_{n}(i)$ & Per-unit cost of resource of type $n$ for node $i\in\mathcal{N}_{\textrm{I}}$\tabularnewline
$c_{\text{b}}(ij)$ & Per-unit cost for link $ij\in\mathcal{E}_{\textrm{I}}$\tabularnewline
$c_{\textrm{f}}(i)$ & Fixed cost for using node $i\in\mathcal{N}_{\textrm{I}}$\tabularnewline
\hline 
\multicolumn{2}{l}{\textbf{SRD graph for slice $\sigma$: }$\mathcal{G}_{\textrm{V}}^{\textrm{\ensuremath{\sigma}}}=\left(\mathcal{N}_{\textrm{V}}^{\textrm{\ensuremath{\sigma}}},\mathcal{E}_{\textrm{V }}^{\textrm{\ensuremath{\sigma}}}\right)$}\tabularnewline
$\mathcal{N}_{\textrm{V}}^{\textrm{\ensuremath{\sigma}}}$ & Set of SRD nodes of slice $\sigma$\tabularnewline
$\mathcal{E}_{\textrm{v}}^{\textrm{\ensuremath{\sigma}}}$ & Set of SRD links of slice $\sigma$\tabularnewline
$v_{\textrm{r}}$ & SRD node aggregating uplink and downlink radio resource demand, $v_{\textrm{r}}\in\mathcal{N}_{\textrm{V}}^{\textrm{\ensuremath{\sigma}}}$\tabularnewline
$r_{n}(v)$ & Resource demand of type $n$ at node $v\in\mathcal{N}_{\textrm{V}}^{\textrm{\ensuremath{\sigma}}}$\tabularnewline
$r_{\text{b}}(vw)$ & Bandwidth demand at link $vw\in\mathcal{E}_{\textrm{V }}^{\textrm{\ensuremath{\sigma}}}$\tabularnewline
$\mathcal{A}^{\textrm{\ensuremath{\sigma}}}$ & Coverage area of slice $\sigma$\tabularnewline
$\mathcal{Q}^{\sigma}$ & Set of all divided subareas in $\mathcal{A}^{\textrm{\ensuremath{\sigma}}}$\tabularnewline
$q$ & Subarea index, $q\in\mathcal{Q}^{\sigma}$\tabularnewline
$\mathcal{A}_{q}^{\sigma}$ & Subarea $q$\tabularnewline
\hline 
$\sigma$ & Slice index\tabularnewline
$\mathcal{S}$ & Set of all slices $\sigma$\tabularnewline
\hline 
\end{tabular}
\end{table}

\section{Problem Formulation\label{sec:ProblemFormulation}}

\begin{figure}[tbh]
\begin{centering}
\includegraphics[width=0.7\columnwidth]{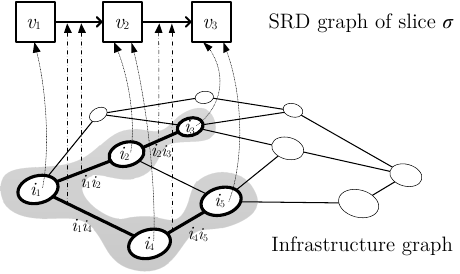}
\par\end{centering}
\caption{Provisioning of infrastructure resource to an SRD graph: Resources
from the infrastructure node $i_{1}$ is provisioned for SRD node
$v_{1}$; Resources from $i_{2}$ and $i_{4}$ are provisioned for
SRD node $v_{2}$; and resources from $i_{3}$ and $i_{5}$ are provisioned
for SRD node $v_{3}$. Correspondingly, the infrastructure links $i_{1}i_{2}$
and $i_{1}i_{4}$ are provisioned for SRD link $v_{1}v_{2}$ and resources
from links $i_{2}i_{3}$ and $i_{4}i_{5}$ are provisioned for SRD
link $v_{2}v_{3}$. \label{fig:SRDProvExample}}
\end{figure}

The provisioning is represented by a mapping between the infrastructure
graph $\mathcal{G}_{\textrm{I}}$ and the SRD graph $\mathcal{G}_{\textrm{V}}^{\sigma}$,
as illustrated in Figure~\ref{fig:SRDProvExample}. In this example,
the slice $\sigma$ is described by an SRD graph aggregating the demands
of several linear SFCs. The constraints that have to be satisfied
by this mapping are detailed in the following sections.

\subsection{Accounting for SRD Coverage Constraints\label{subsec:Formulation_ProvRadio}}

For the slice $\sigma$, the InP has to provide a minimum average
data rate ($\underline{R}_{\textrm{u}}^{\sigma}$ for uplink and $\underline{R}_{\textrm{d}}^{\sigma}$
for downlink) to each mobile user spread over $\mathcal{A}^{\sigma}$
with a density $\rho^{\sigma}\left(x\right)$. For that purpose, the
InP will have to provision resources from the physical RRH nodes in
$\mathcal{N}_{\text{Ir}}$. One assumes that every RRH node is able
to provide a fixed amount $a_{\text{r}}\left(i\right)$ of resource
blocks (RB) per time unit to exchange data (up and downlink) with
users. The amount of data transmitted using a single RB depends on
the characteristics of the RRH, of the User Equipment (UE), and on
the transmission channel between the RRH and the user.

During the resource provisioning phase, the locations of users are
unknown. To address this problem, \cite{Teague} considers different
realizations of a point process representing the location of users.
Here an approach inspired by the subarea partitioning technique introduced
in \cite{Shi2007} is considered. $\mathcal{A^{\sigma}}$ is partitioned
into $Q^{\sigma}$ convex subareas $\mathcal{A}_{q}^{\sigma}$, $q\in\mathcal{Q}^{\sigma}=\left\{ 1,\dots,Q^{\sigma}\right\} $.
Instead of allocating RBs to users, RRH nodes allocate RBs to subareas.
The way the partitioning is performed is not detailed here. One may
consider, \emph{e.g.}, a partitioning into squares of equal surfaces
or a partitioning based on $\rho^{\sigma}$ that provides an equal
average number of users per subarea.

For slice~$\sigma$, the proportion of RBs provisioned by RRH~$i$
to the users in $\mathcal{A}_{q}^{\sigma}$ is denoted by $\eta_{\text{u}}^{\sigma}\left(i,q\right)\in\left[0,1\right]$
and $\eta_{\text{d}}^{\sigma}\left(i,q\right)\in\left[0,1\right]$
for uplink and downlink traffic, respectively. These quantities represent
average proportions of RBs available during some typical interval
of time and provisioned by RRH~$i$. The time interval may be, \emph{e.g.},
of one second\footnote{Since $\eta_{\text{u}}^{\sigma}\left(i,q\right)$ and $\eta_{\text{d}}^{\sigma}\left(i,q\right)$
are averages, they may be accurately represented by real numbers in
the interval $\left[0,1\right]$, even if in reality both quantities
should be rational numbers.}. The summed proportions of RBs provided by a given RRH~$i$ must
be less than one
\begin{equation}
\sum_{\sigma\in\mathcal{S}}\sum_{q\in\mathcal{Q}^{\sigma}}\left(\eta_{\text{u}}^{\sigma}\left(i,q\right)+\eta_{\text{d}}^{\sigma}\left(i,q\right)\right)\leqslant1,\,\forall i\in\mathcal{N}_{\text{Ir}}.\label{eq:Cover_C1_EtaLessThanOne}
\end{equation}

For each slice $\sigma$ and each subarea $\mathcal{A}_{q}^{\sigma}$,
the total data rate provided by the allocated resource blocks should
satisfy the minimum average user demand. Then, $\forall q\in\mathcal{Q}^{\sigma},\forall\sigma\in\mathcal{S}$,
one should have
\begin{align}
 & \sum_{i\in\mathcal{N}_{\text{Ir}}}\eta_{\text{u}}^{\sigma}\left(i,q\right)a_{\text{r}}\left(i\right)b_{\text{u}}\left(x_{i}^{\textrm{r}},\mathcal{A}_{q}^{\sigma}\right)\geqslant\underline{R}_{\text{u}}^{\sigma}\int_{\mathcal{A}_{q}^{\sigma}}\rho^{\sigma}\left(x\right)\mathrm{d}x,\label{eq:Cover_C2_RequiredMinRateUL}\\
 & \sum_{i\in\mathcal{N}_{\text{Ir}}}\eta_{\text{d}}^{\sigma}\left(i,q\right)a_{\text{r}}\left(i\right)b_{\text{d}}\left(x_{i}^{\textrm{r}},\mathcal{A}_{q}^{\sigma}\right)\geqslant\underline{R}_{\text{d}}^{\sigma}\int_{\mathcal{A}_{q}^{\sigma}}\rho^{\sigma}\left(x\right)\mathrm{d}x,\label{eq:Cover_C3_RequiredMinRateDL}
\end{align}
which correspond to the satisfaction of the geographical coverage
constraints for uplink and downlink traffic. Here, $b_{\text{u}}\left(x_{i}^{\textrm{r}},\mathcal{A}_{q}^{\sigma}\right)$
and $b_{\text{d}}\left(x_{i}^{\textrm{r}},\mathcal{A}_{q}^{\sigma}\right)$
denote the amount of data (bits) carried by a RB for a user located
in $\mathcal{A}_{q}^{\sigma}$ for up and downlink. Depending on the
level of conservatism, $b_{\text{u}}\left(x_{i}^{\textrm{r}},\mathcal{A}_{q}^{\sigma}\right)$
and $b_{\text{d}}\left(x_{i}^{\textrm{r}},\mathcal{A}_{q}^{\sigma}\right)$
may represent the minimum or the average amount of data evaluated
over the possible locations of users in $\mathcal{A}_{q}^{\sigma}$.
The terms $b_{\text{u}}\left(x_{i}^{\textrm{r}},\mathcal{A}_{q}^{\sigma}\right)$,
$b_{\text{d}}\left(x_{i}^{\textrm{r}},\mathcal{A}_{q}^{\sigma}\right)$,
and $\int_{\mathcal{A}_{q}^{\sigma}}\rho^{\sigma}\left(x\right)\mathrm{d}x$
are fixed quantities that only depend on the RRH location $x_{i}^{\textrm{r}}$,
on the user density $\rho^{\sigma}$, and on the way the partitioning
of $\mathcal{A}^{\sigma}$ has been performed. These terms may thus
be evaluated in advance, see Section~\ref{subsec:Rate-Allocation-Model}.
Summing \eqref{eq:Cover_C2_RequiredMinRateUL} over all $q\in\mathcal{Q}^{\sigma}$
and using \eqref{eq:SRDUpDownlink}, one gets
\begin{align}
 & \sum_{q\in\mathcal{Q}^{\sigma}}\sum_{i\in\mathcal{N}_{\text{Ir}}}\eta_{\text{u}}^{\sigma}\left(i,q\right)a_{\text{r}}\left(i\right)b_{\text{u}}\left(x_{i}^{\textrm{r}},\mathcal{A}_{q}^{\sigma}\right)\geqslant r_{\text{u}}^{\sigma}\left(v_{\text{r}}\right),\label{eq:Cover_C4_RadioNodeDemandUL}\\
 & \sum_{q\in\mathcal{Q}^{\sigma}}\sum_{i\in\mathcal{N}_{\text{Ir}}}\eta_{\text{d}}^{\sigma}\left(i,q\right)a_{\text{r}}\left(i\right)b_{\text{d}}\left(x_{i}^{\textrm{r}},\mathcal{A}_{q}^{\sigma}\right)\geqslant r_{\text{d}}^{\sigma}\left(v_{\text{r}}\right),\label{eq:Cover_C5_RadioNodeDemandDL}
\end{align}
which ensure, for slice $\sigma$, the satisfaction of the part of
the SRD graph related to the uplink and downlink radio resource demands.

For each RRH~$i$, the amount of provisioned uplink and downlink
resources should be proportional to the demand expressed in the SRD
graph through $r_{\text{u}}^{\sigma}\left(v_{\text{r}}\right)$ and
$r_{\text{d}}^{\sigma}\left(v_{\text{r}}\right)$. This avoids provisioning
RRH resources taking care only of the uplink or only of the downlink
traffic. This has to be ensured for all subareas $q\in\mathcal{Q}^{\sigma}$
\begin{equation}
\frac{\eta_{\text{u}}^{\sigma}\left(i,q\right)a_{\text{r}}\left(i\right)b_{\text{u}}\left(x_{i}^{\textrm{r}},\mathcal{A}_{q}^{\sigma}\right)}{r_{\text{u}}^{\sigma}\left(v_{\text{r}}\right)}=\frac{\eta_{\text{d}}^{\sigma}\left(i,q\right)a_{\text{r}}\left(i\right)b_{\text{d}}\left(x_{i}^{\textrm{r}},\mathcal{A}_{q}^{\sigma}\right)}{r_{\text{d}}^{\sigma}\left(v_{\text{r}}\right)}.\label{eq:Cover_C6_UpDownProportionality}
\end{equation}

To identify whether a RRH~$i\in\mathcal{N}_{\textrm{Ir}}$ has provisioned
some RBs to any subarea for slice $\sigma$, one introduces the variables
$\widetilde{\eta}^{\sigma}\left(i\right)\in\left\{ 0,1\right\} $,
with $\widetilde{\eta}^{\sigma}\left(i\right)=1$ if $\sum_{q\in\mathcal{Q}^{\sigma}}\eta^{\sigma}\left(i,q\right)>0$,
and $\widetilde{\eta}^{\sigma}\left(i\right)=0$ otherwise. The variables
$\eta^{\sigma}\left(i,q\right)$ and $\widetilde{\eta}^{\sigma}\left(i\right)$
are gathered in the sets $\boldsymbol{\eta}^{\sigma}=\{\eta^{\sigma}\left(i,q\right)\}_{i\in\mathcal{\mathcal{N}_{\text{Ir}}},q\in\mathcal{Q}^{\sigma}}$
and $\widetilde{\boldsymbol{\eta}}^{\sigma}=\{\widetilde{\eta}^{\sigma}\left(i\right)\}_{i\in\mathcal{\mathcal{N}_{\text{Ir}}}},$
which are components of the sets $\boldsymbol{\eta}=\{\boldsymbol{\eta}^{\sigma}\}_{\sigma\in\mathcal{S}}$
and $\widetilde{\boldsymbol{\eta}}=\{\widetilde{\boldsymbol{\eta}}^{\sigma}\}_{\sigma\in\mathcal{S}}$.
The relation between $\eta^{\sigma}\left(i,q\right)$ and $\widetilde{\eta}^{\sigma}\left(i\right)$
is nonlinear. Nevertheless, both quantities can be linked with the
following linear constraints, $\forall\sigma\in\mathcal{S}$, $\forall i\in\mathcal{N}_{\text{Ir}},$
\begin{equation}
0\leq\widetilde{\eta}^{\sigma}\left(i\right)-\sum_{q\in\mathcal{Q}^{\sigma}}\eta^{\sigma}\left(i,q\right)<1,\label{eq:Cover_C7_RelationEta}
\end{equation}
with
\begin{equation}
\eta^{\sigma}\left(i,q\right)=\eta_{\text{u}}^{\sigma}\left(i,q\right)+\eta_{\text{d}}^{\sigma}\left(i,q\right).\label{eq:Cover_C8_EtaUpDown}
\end{equation}

The leasing cost related to the radio resource provisioning for a
given slice $\sigma$ gathers the fixed costs $c_{\textrm{f}}\left(i\right)\widetilde{\eta}^{\sigma}\left(i\right)$
related to the use of a RRH by the slice and the variable costs $c_{\textrm{r}}\left(i\right)a_{\text{r}}\left(i\right)\eta^{\sigma}\left(i,q\right)$
related to the amount of RBs provided by each RRH to the slice. A
bias towards RB allocation by RRHs providing a high spectral efficiency
is obtained by the introduction of a rate-related discount $\lambda b\left(x_{i}^{\textrm{r}},\mathcal{A}_{q}^{\sigma}\right)a_{\text{r}}\left(i\right)\eta^{\sigma}\left(i,q\right)$,
where $\lambda$ is a positive discount factor. The resulting cost
function for the \emph{radio} resources is
\begin{equation}
\begin{split}c_{\textrm{rr}} & \left(\boldsymbol{\eta},\widetilde{\boldsymbol{\eta}}\right)=\sum_{\sigma\in\mathcal{S}}c_{\textrm{rr}}^{\sigma}\left(\boldsymbol{\eta}^{\sigma},\widetilde{\boldsymbol{\eta}}^{\sigma}\right)\end{split}
,\label{eq:Cover_Cost}
\end{equation}
where
\begin{equation}
\begin{split}c_{\textrm{rr}}^{\sigma} & \left(\boldsymbol{\eta}^{\sigma},\widetilde{\boldsymbol{\eta}}^{\sigma}\right)=\sum_{i\in\mathcal{N}_{\text{Ir}}}c_{\textrm{f}}\left(i\right)\widetilde{\eta}^{\sigma}\left(i\right)\\
+ & \sum_{\sigma\in\mathcal{S}}\sum_{i\in\mathcal{N}_{\text{Ir}}}\sum_{q\in\mathcal{Q}^{\sigma}}\left[c_{\textrm{r}}\left(i\right)-\lambda b_{\text{u}}\left(x_{i}^{\textrm{r}},\mathcal{A}_{q}^{\sigma}\right)\right]a_{\text{r}}\left(i\right)\eta_{\text{u}}^{\sigma}\left(i,q\right)\\
+ & \sum_{\sigma\in\mathcal{S}}\sum_{i\in\mathcal{N}_{\text{Ir}}}\sum_{q\in\mathcal{Q}^{\sigma}}\left[c_{\textrm{r}}\left(i\right)-\lambda b_{\text{d}}\left(x_{i}^{\textrm{r}},\mathcal{A}_{q}^{\sigma}\right)\right]a_{\text{r}}\left(i\right)\eta_{\text{d}}^{\sigma}\left(i,q\right)
\end{split}
\label{eq:Cover_Cost_Sigma}
\end{equation}

\subsection{Accounting for other SRD Constraints\label{subsec:Formulation_ProvOther}}

This section introduces a set of constraints which have to be satisfied
to address the other resource demands for each $\text{\ensuremath{\sigma}\ensuremath{\in}}\mathcal{S}$,
while being consistent with the coverage constraints.

For that purpose, one introduces first $\boldsymbol{\Phi}_{\textrm{n}}^{\sigma}=\left\{ \phi_{n}^{\sigma}(i,v)\right\} _{i\in\mathcal{N}_{\textrm{I}},v\in\mathcal{N}_{\textrm{V}}^{\textrm{\ensuremath{\sigma}}},n\in\{\textrm{c},\textrm{s}\}}$,
where $\phi_{n}^{\sigma}(i,v)$ represents the proportion of resources
of type $n\in\{\textrm{c},\textrm{s}\}$ provisioned on the infrastructure
node $i\in\mathcal{G}_{\textrm{I}}$ for the SRD node $v\in\mathcal{N}_{\textrm{V}}^{\textrm{\ensuremath{\sigma}}}$
of slice $\sigma$. Second, let $\boldsymbol{\Phi}_{\text{b}}^{\sigma}=\left\{ \phi_{\text{b}}^{\sigma}(ij,vw)\right\} _{ij\in\mathcal{E}_{\textrm{I}},vw\in\mathcal{E}_{\textrm{V}}^{\textrm{\ensuremath{\sigma}}}}$,
where $\phi_{\text{b}}^{\sigma}(ij,vw)$ represents the proportion
of bandwidth of the infrastructure link $ij\in\mathcal{E}_{\textrm{I}}$
provisioned for the SRD link $vw\in\mathcal{E}_{\textrm{V}}^{\textrm{\ensuremath{\sigma}}}$
of slice $\sigma$. The sets $\boldsymbol{\Phi}_{\textrm{n}}=\left\{ \boldsymbol{\Phi}_{\textrm{n}}^{\sigma}\right\} _{\sigma\in\mathcal{S}}$
and $\boldsymbol{\Phi}_{\textrm{b}}=\left\{ \boldsymbol{\Phi}_{\textrm{b}}^{\sigma}\right\} _{\sigma\in\mathcal{S}}$
are sets of non-negative real variables ranging from $0$ to $1$.
When one of the variables holds zero, there is no mapping between
the infrastructure and the SRD node/link.

The sum of resources provided by each infrastructure node $i\in\mathcal{N_{\textrm{I}}}$
mapped to an SRD node $v$ should satisfy its resource demands. This
leads $\forall\sigma\in\mathcal{S}$ to
\begin{equation}
\sum\limits _{i\in\mathcal{N_{\textrm{I}}}}a_{n}\left(i\right)\phi_{n}^{\sigma}(i,v)\geq r_{n}(v),\,\forall n\in\{\textrm{c},\textrm{s}\},\forall v\in\mathcal{N}_{\textrm{V}}^{\sigma}.\label{eq:Prov_C1_Node_Satisfied}
\end{equation}
Since, the summed proportions of resources provisioned by a given
infrastructure node $i$ cannot exceed one, we have
\begin{equation}
\sum_{\sigma\in\mathcal{S}}\sum\limits _{v\in\mathcal{N}_{\textrm{V}}^{\textrm{\ensuremath{\sigma}}}}\phi_{n}^{\sigma}(i,v)\leq1,\,\forall n\in\{\textrm{c},\textrm{s}\},\forall i\in\mathcal{N}_{\textrm{I}}.\label{eq:Prov_C2_Node_Limit}
\end{equation}
Similarly, the cumulative proportions of resources provisioned by
a given infrastructure link $ij$ cannot exceed one
\begin{equation}
\sum_{\sigma\in\mathcal{S}}\sum\limits _{vw\in\mathcal{E}_{\textrm{V }}^{\textrm{\ensuremath{\sigma}}}}\phi_{\text{b}}^{\sigma}(ij,vw)\leq1,\,\forall ij\in\mathcal{E}_{\textrm{I}}.\label{eq:Prov_C3_Link_Limit}
\end{equation}

The amount of resources provided by a given infrastructure node $i$
to an SRD node $v$ has to be equal to an integer multiple of the
minimum amount of resources $\underline{r}_{n}\left(v\right)$ for
a VNF associated to the SRD node~$v$
\begin{equation}
\begin{split} & a_{n}(i)\phi_{n}^{\sigma}(i,v)=\underline{r}_{n}(v)\kappa_{n}^{\sigma}(i,v),\\
 & \forall i\in\mathcal{N}_{\textrm{I}},\forall v\in\mathcal{N}_{\textrm{V}}^{\textrm{\ensuremath{\sigma}}},\forall n\in\{\textrm{c},\textrm{s}\},
\end{split}
\label{eq:Prov_C4_Node_Min_Requirement}
\end{equation}
where $\kappa_{n}^{\sigma}(i,v)$ is a positive integer belonging
to the set of variables of the optimization problem. This ensures
that enough resources are provisioned by an infrastructure node $i$
to be able to deploy an integer number $\kappa_{n}^{\sigma}(i,v)$
of VNF instances associated to the SRD node $v$.

It is usually difficult, if not impossible, when deploying a given
VNF, to benefit from the storage of one infrastructure node and from
the computing resources of another infrastructure node. Consequently,
resources of each type have to be provisioned in a balanced way by
an infrastructure node for an SRD node, consistently with the requirements
of the SRD node. This ensures to be able to deploy a VNF on a single
infrastructure node. For example, if an infrastructure node provides
$10\%$ of the computing demand of a given SRD node, it should also
provide $10\%$ of its storage demand. This translates into the following
resource provisioning proportionality constraints $\forall\sigma\in\mathcal{S}$,
\begin{equation}
\cfrac{a_{\textrm{c}}(i)}{r_{\textrm{c}}(v)}\phi_{\text{c}}^{\sigma}(i,v)=\cfrac{a_{\textrm{s}}(i)}{r_{\textrm{s}}(v)}\phi_{\textrm{s}}^{\sigma}(i,v),\text{ }\forall i\in\mathcal{N}_{\textrm{I}},\forall v\in\mathcal{N}_{\textrm{V}}^{\textrm{\ensuremath{\sigma}}}.\label{eq:Prov_C5_Proportionality}
\end{equation}

Additionally, considering the SRD node $v_{\text{r}}$, the computing
and storage resources provisioned by an infrastructure node $i\in\mathcal{N}_{\textrm{Ir}}$
should be commensurate with the provisioned wireless resources, $\forall\sigma\in\mathcal{S}$
and $\forall i\in\mathcal{N}_{\textrm{Ir}}$,
\begin{equation}
\begin{split} & \cfrac{a_{\textrm{c}}(i)}{r_{\textrm{c}}(v_{\text{r}})}\phi_{\text{c}}^{\sigma}(i,v_{\text{r}})=\cfrac{a_{\textrm{s}}(i)}{r_{\textrm{s}}(v_{\text{r}})}\phi_{\textrm{s}}^{\sigma}(i,v_{\text{r}})=\frac{a_{\text{r}}\left(i\right)}{r_{\text{r}}^{\sigma}(v_{\text{r}})}\\
 & \times\sum_{q\in\mathcal{Q}^{\sigma}}\left(\eta_{\text{u}}^{\sigma}\left(i,q\right)b_{\text{u}}\left(x_{i}^{\textrm{r}},\mathcal{A}_{q}^{\sigma}\right)+\eta_{\text{d}}^{\sigma}\left(i,q\right)b_{\text{d}}\left(x_{i}^{\textrm{r}},\mathcal{A}_{q}^{\sigma}\right)\right).
\end{split}
\label{eq:Prov_C5_ProportionalityRadio}
\end{equation}
The constraints \eqref{eq:Prov_C5_Proportionality} and \eqref{eq:Prov_C5_ProportionalityRadio}
ensure a balanced resource provisioning by infrastructure nodes. In
\eqref{eq:Prov_C5_ProportionalityRadio}, $r_{\text{r}}^{\sigma}(v_{\text{r}})$
is the total radio resource demand of $v_{\textrm{r}}$ in both up
and downlink, \textit{i.e.}, $r_{\text{r}}^{\sigma}(v_{\text{r}})=r_{\text{u}}^{\sigma}(v_{\text{r}})+r_{\text{d}}^{\sigma}(v_{\text{r}})$.

Moreover, link resources should be consistently provisioned with the
radio resource of the RRH for both uplink and downlink. Thus, for
downlink traffic (links with RRH as egress), one should have $\forall\sigma\in\mathcal{S}$,
$\forall j\in\mathcal{N}_{\textrm{Ir}}$, $\forall vv_{\text{r}}\in\mathcal{E}_{\textrm{V }}^{\textrm{\ensuremath{\sigma}}}$,
\begin{align}
\sum_{i\in\mathcal{N}_{\textrm{I}}\backslash\mathcal{N}_{\textrm{Ir}}}\cfrac{a_{\text{b}}(ij)}{r_{\text{b}}(vv_{\text{r}})}\phi_{\text{b}}^{\sigma}(ij,vv_{\text{r}})= & \left(\cfrac{r_{\text{b}}(vv_{\text{r}})}{\sum_{uv_{\text{r}}\in\mathcal{E}_{\textrm{V}}^{\textrm{\ensuremath{\sigma}}}}r_{\text{b}}(uv_{\text{r}})}\right)\frac{a_{\text{r}}\left(j\right)}{r_{\text{d}}^{\sigma}(v_{\text{r}})}\nonumber \\
\times & \sum_{q\in\mathcal{Q}^{\sigma}}\eta_{\text{d}}^{\sigma}\left(j,q\right)b_{\text{d}}\left(x_{j}^{\textrm{r}},\mathcal{A}_{q}^{\sigma}\right)\label{eq:Prov_RadioDownlink}
\end{align}
In \eqref{eq:Prov_RadioDownlink}, the term $a_{\text{r}}\left(j\right)\sum_{q\in\mathcal{Q}^{\sigma}}\eta_{\text{d}}^{\sigma}\left(j,q\right)b_{\text{d}}\left(x_{j}^{\textrm{r}},\mathcal{A}_{q}^{\sigma}\right)/r_{\text{d}}^{\sigma}(v_{\text{r}})$
represents the proportion of downlink radio resources provided by
RRH~$j$ to satisfy the downlink demand of $v_{\text{r}}$. When
several SRD links feed $v_{\text{r}}$, the term $r_{\text{b}}(vv_{\text{r}})/\sum_{uv_{\text{r}}\in\mathcal{E}_{\textrm{V}}^{\textrm{\ensuremath{\sigma}}}}r_{\text{b}}(uv_{\text{r}})$
represents the proportion of (downlink) traffic demand associated
to the SRD link~$vv_{\text{r}}$. The right-hand side of \eqref{eq:Prov_RadioDownlink}
represents thus the proportion of the data traffic that \emph{has
to be provisioned} for the SRD link~$vv_{\text{r}}$ to satisfy the
part of the downlink radio resource provided by RRH~$j$ to satisfy
the part of the downlink demand of $v_{\text{r}}$. The left-hand
side of \eqref{eq:Prov_RadioDownlink}, represents the proportion
of the data traffic that \emph{is provided} by all infrastructure
links $ij$, $i\in\mathcal{N}_{\textrm{I}}\backslash\mathcal{N}_{\textrm{Ir}}$
for the SRD link $vv_{\text{r}}$. Both terms have thus to be equal.

For uplink traffic (links with RRH as ingress), one has, $\forall\sigma\in\mathcal{S}$,
$\forall i\in\mathcal{N}_{\textrm{Ir}}$, $\forall v_{\text{r}}v\in\mathcal{E}_{\textrm{V }}^{\textrm{\ensuremath{\sigma}}}$,
\begin{align}
\sum_{j\in\mathcal{N}_{\textrm{I}}\backslash\mathcal{N}_{\textrm{Ir}}}\cfrac{a_{\text{b}}(ij)}{r_{\text{b}}(v_{\text{r}}v)}\phi_{\text{b}}^{\sigma}(ij,v_{\text{r}}v)= & \left(\cfrac{r_{\text{b}}(v_{\text{r}}v)}{\sum_{v_{\text{r}}u\in\mathcal{E}_{\textrm{V}}^{\textrm{\ensuremath{\sigma}}}}r_{\text{b}}(v_{\text{r}}u)}\right)\frac{a_{\text{r}}\left(i\right)}{r_{\text{u}}^{\sigma}(v_{\text{r}})}\nonumber \\
\times & \sum_{q\in\mathcal{Q}^{\sigma}}\eta_{\text{u}}^{\sigma}\left(i,q\right)b_{\text{u}}\left(x_{i}^{\textrm{r}},\mathcal{A}_{q}^{\sigma}\right)\label{eq:Prov_RadioUplink}
\end{align}
In \eqref{eq:Prov_RadioUplink}, the term $a_{\text{r}}\left(i\right)\sum_{q\in\mathcal{Q}^{\sigma}}\eta_{\text{u}}^{\sigma}\left(i,q\right)b_{\text{u}}\left(x_{i}^{\textrm{r}},\mathcal{A}_{q}^{\sigma}\right)/r_{\text{u}}^{\sigma}(v_{\text{r}})$
represents now the proportion of uplink radio resources provided by
RRH~$i$ to satisfy the uplink demand of $v_{\text{r}}$. When several
SRD links depart from $v_{\text{r}}$, the term $r_{\text{b}}(v_{\text{r}}v)/\sum_{v_{\text{r}}u\in\mathcal{E}_{\textrm{V}}^{\textrm{\ensuremath{\sigma}}}}r_{\text{b}}(v_{\text{r}}u)$
represents the proportion of (uplink) traffic demand associated to
the SRD link~$v_{\text{r}}v$. The right-hand side of \eqref{eq:Prov_RadioUplink}
represents thus the proportion of the data traffic that \emph{has
to be provisioned} for the SRD link~$v_{\text{r}}v$ to convey the
part of the uplink radio resource provided by RRH~$i$ to satisfy
the part of the uplink demand of $v_{\text{r}}$. The left-hand side
of \eqref{eq:Prov_RadioUplink}, represents the proportion of the
data traffic that \emph{is provided} by all infrastructure links $ij$,
$j\in\mathcal{N}_{\textrm{I}}\backslash\mathcal{N}_{\textrm{Ir}}$
for the SRD link $v_{\text{r}}v$. Both terms have again to be equal.
Combined with \eqref{eq:Cover_C5_RadioNodeDemandDL}, the constraints
\eqref{eq:Prov_RadioDownlink} and \eqref{eq:Prov_RadioUplink} impose
that the total radio resources provisioned by the RRHs are above the
required resources $r_{\text{d}}^{\sigma}\left(v_{\text{r}}\right)$
and $r_{\text{u}}^{\sigma}\left(v_{\text{r}}\right)$.

Finally, flow conservation constraints have to be satisfied when resources
are provisioned on the infrastructure link $ij$ for the SRD link
$vw$. That is, for each SRD link $vw\in\mathcal{E}_{\textrm{v }}$,
a path of infrastructure links must be provisioned between \textit{each}
pair of infrastructure nodes that are mapped to the pair $\left(v,w\right)$
of SRD nodes.

Consider first an infrastructure node $i$ which provisions resources
for two SRD nodes $v$ and $w$. The corresponding VNFs will have
to exchange information within the considered node~$i$ via the internal
link $ii$. For such internal link providing resources to an SRD link,
one should have $\forall\sigma\in\mathcal{S}$, $\forall i\in\mathcal{N}_{\text{I}}$,
$\forall vw\in\mathcal{E}_{\textrm{V }}^{\textrm{\ensuremath{\sigma}}}$,
\begin{align}
\cfrac{a_{\text{b}}(ii)}{r_{\text{b}}(vw)}\phi_{\text{b}}^{\sigma}(ii,vw) & =\left(\frac{r_{\text{b}}(vw)}{\sum_{vu\in\mathcal{E}_{\textrm{V}}^{\textrm{\ensuremath{\sigma}}}}r_{\text{b}}(vu)}\right)\cfrac{a_{\textrm{c}}(i)}{r_{\textrm{c}}(v)}\phi_{\textrm{c}}^{\sigma}(i,v)\label{eq:Prov_C7_Internal_Link1}\\
 & =\left(\frac{r_{\text{b}}(vw)}{\sum_{uw\in\mathcal{E}_{\textrm{V}}^{\textrm{\ensuremath{\sigma}}}}r_{\text{b}}(uw)}\right)\cfrac{a_{\textrm{c}}(i)}{r_{\textrm{c}}(w)}\phi_{\textrm{c}}^{\sigma}(i,w),\label{eq:Prov_C7_Internal_Link2}
\end{align}
In \eqref{eq:Prov_C7_Internal_Link1}, the term $a_{\textrm{c}}(i)\phi_{\textrm{c}}^{\sigma}(i,v)/r_{\textrm{c}}(v)$
represents the proportion of computing resource provided by the infrastructure
node~$i$ to meet the demand of the SRD node~$v$. When several
SRD links depart from $v$, the term $r_{\text{b}}(vw)/\sum_{vu\in\mathcal{E}_{\textrm{V}}^{\textrm{\ensuremath{\sigma}}}}r_{\text{b}}(vu)$
represents the proportion of traffic demand that departs from $v$
associated to the SRD link~$vw$. The right-hand side of \eqref{eq:Prov_C7_Internal_Link1}
represents thus the proportion of the data traffic that \emph{has
to be provisioned} for the SRD link~$vw$ to satisfy the corresponding
proportion of computing resources provided by~$i$ to satisfy the
part of the demand of $v$. The left-hand side of \eqref{eq:Prov_RadioUplink}
represents the proportion of the data traffic that \emph{is provided}
by the internal link $ii$ for the SRD link $vw$. The constraint
\eqref{eq:Prov_C7_Internal_Link2} can be justified similarly.

Consider now an SRD link~$vw$ and an infrastructure node $i$ which
provisions resources either for only one of the SRD nodes $v$ or
$w$, or for none of them. Focusing again on the computing resource,
three cases have to be considered.

Assume first that $i$ provisions resources for $v$. Then, one has
the following constraint $\forall\sigma\in\mathcal{S}$
\begin{equation}
\sum\limits _{j\in\mathcal{N}_{\textrm{I}}}\cfrac{a_{\text{b}}(ij)}{r_{\text{b}}(vw)}\phi_{\text{b}}^{\sigma}(ij,vw)=\left(\frac{r_{\text{b}}(vw)}{\sum_{vu\in\mathcal{E}_{\textrm{V}}^{\textrm{\ensuremath{\sigma}}}}r_{\text{b}}(vu)}\right)\cfrac{a_{\textrm{c}}(i)}{r_{\textrm{c}}(v)}\phi_{\textrm{c}}^{\sigma}(i,v),\label{eq:Flow_Conservation1}
\end{equation}
The right-hand side of \eqref{eq:Flow_Conservation1} is the same
as that of \eqref{eq:Prov_C7_Internal_Link1}. The left-hand side
of \eqref{eq:Flow_Conservation1} represents the proportion of the
traffic provisioned by all links~$ij$, $j\in\mathcal{N}_{\textrm{I}}$
(leaving node~$i$) for the SRD link~$vw$.

Assume second that $i$ provisions resources for $w$. Then, one has
the following constraint $\forall\sigma\in\mathcal{S}$
\begin{equation}
\sum\limits _{j\in\mathcal{N}_{\textrm{I}}}\cfrac{a_{\text{b}}(ij)}{r_{\text{b}}(vw)}\phi_{\text{b}}^{\sigma}(ji,vw)=\left(\frac{r_{\text{b}}(vw)}{\sum_{uw\in\mathcal{E}_{\textrm{V}}^{\textrm{\ensuremath{\sigma}}}}r_{\text{b}}(uw)}\right)\cfrac{a_{\textrm{c}}(i)}{r_{\textrm{c}}(w)}\phi_{\textrm{c}}^{\sigma}(i,w)\label{eq:Flow_Conservation2}
\end{equation}
The right-hand side of \eqref{eq:Flow_Conservation2} is the same
as that of \eqref{eq:Prov_C7_Internal_Link2}. The left-hand side
of \eqref{eq:Flow_Conservation2} represents now the proportion of
the traffic provisioned by all links~$ji$, $j\in\mathcal{N}_{\textrm{I}}$
(feeding node $i$) for the SRD link~$vw$.

Assume finally that $i$ provisions resources neither for $v$ nor
for $w$. Then, the following flow-conservation constraint
\begin{equation}
\sum\limits _{j\in\mathcal{N}_{\textrm{I}}}\left[\cfrac{a_{\text{b}}(ij)}{r_{\text{b}}(vw)}\phi_{\text{b}}^{\sigma}(ij,vw)-\cfrac{a_{\text{b}}(ji)}{r_{\text{b}}(vw)}\phi_{\text{b}}^{\sigma}(ji,vw)\right]=0\label{eq:Flow_Conservation3}
\end{equation}
must be satisfied $\forall\sigma\in\mathcal{S}$.

The constraints (\ref{eq:Flow_Conservation1}-\ref{eq:Flow_Conservation3})
can be gathered in the following single constraint, which should be
valid, $\forall\sigma\in\mathcal{S}$, $\forall i\in\mathcal{N}_{\textrm{I}}$,
$\forall vw\in\mathcal{E}_{\textrm{V }}^{\textrm{\ensuremath{\sigma}}}$,
\begin{align}
 & \sum\limits _{j\in\mathcal{N}_{\textrm{I}}}\left[\cfrac{a_{\text{b}}(ij)}{r_{\text{b}}(vw)}\phi_{\text{b}}^{\sigma}(ij,vw)-\cfrac{a_{\text{b}}(ji)}{r_{\text{b}}(vw)}\phi_{\text{b}}^{\sigma}(ji,vw)\right]\nonumber \\
 & =\left(\frac{r_{\text{b}}(vw)}{\sum_{vu\in\mathcal{E}_{\textrm{V}}^{\textrm{\ensuremath{\sigma}}}}r_{\text{b}}(vu)}\right)\cfrac{a_{\textrm{c}}(i)}{r_{\textrm{c}}(v)}\phi_{\textrm{c}}^{\sigma}(i,v)\nonumber \\
 & -\left(\frac{r_{\text{b}}(vw)}{\sum_{uw\in\mathcal{E}_{\textrm{V}}^{\textrm{\ensuremath{\sigma}}}}r_{\text{b}}(uw)}\right)\cfrac{a_{\textrm{c}}(i)}{r_{\textrm{c}}(w)}\phi_{\textrm{c}}^{\sigma}(i,w),\label{eq:Prov_C6_FlowConservation}
\end{align}

In \eqref{eq:Prov_C7_Internal_Link1}, \eqref{eq:Prov_C7_Internal_Link2},
and \eqref{eq:Prov_C6_FlowConservation}, the consistency with the
other provisioned resources is ensured by \eqref{eq:Prov_C5_Proportionality}.

Note that the flow conservation constraints~\eqref{eq:Prov_C6_FlowConservation}
imposes a relation between the $\phi_{n}^{\sigma}(i,v)$s for different
$i$ and $v$. Since $\phi_{n}^{\sigma}(i,v)$ and $\kappa_{n}^{\sigma}(i,v)$
are proportional, see \eqref{eq:Prov_C4_Node_Min_Requirement}, the
relations between $\kappa_{n}^{\sigma}(i,v)$ for different $i$ and
$v$ are also imposed without specifying any additional constraint.

\begin{figure}[t]
\begin{centering}
\subfloat[Fork topology\label{fig:Example_Split}]{\begin{centering}
\includegraphics[width=0.4\columnwidth]{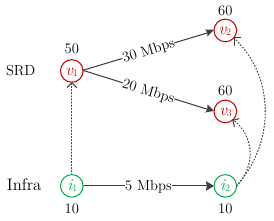}
\par\end{centering}
}\subfloat[Merge topology\label{fig:Example_Joint}]{\begin{centering}
\includegraphics[width=0.4\columnwidth]{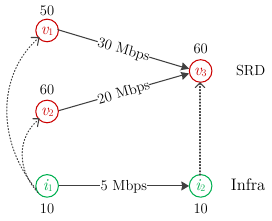}
\par\end{centering}
}
\par\end{centering}
\caption{Illustration to the constraint~\eqref{eq:Prov_C6_FlowConservation}
related to flow conservation considering an SRD graph with (a) a fork
topology and (b) merge topology. The nodes $i_{1},i_{2}$ and the
link $i_{1}i_{2}$ belong to the infrastructure graph; The nodes $v_{1},v_{2},v_{3}$
and the links connecting them belong to the SRD graph. \label{fig:Example:Coefficient}}
\end{figure}

Figure~\ref{fig:Example:Coefficient} illustrates two resource provisioning
examples for simplified SRD graphs with branched topologies. In Figure~\ref{fig:Example_Split},
the node $i_{1}$ is mapped onto $v_{1}$, $i_{2}$ is mapped onto
the pair $\left(v_{2},v_{3}\right)$; and the link $i_{1}i_{2}$ is
mapped onto the pair $\left(v_{1}v_{2},v_{1}v_{3}\right)$. Considering
the infrastructure node $i_{1}$ and the SRD link $v_{1}v_{2},$ the
constraint \eqref{eq:Prov_C6_FlowConservation} leads to $\frac{5}{30}\phi_{\text{b}}\left(i_{1}i_{2},v_{1}v_{2}\right)-0=\frac{30}{50}\frac{10}{50}\phi_{n}\left(i_{1},v_{1}\right)-0$,
hence $\phi_{\text{b}}\left(i_{1}i_{2},v_{1}v_{2}\right)=\frac{18}{25}\phi_{n}\left(i_{1},v_{1}\right)$.
Similarly, considering $i_{1}$ and $v_{1}v_{3}$, one gets $\phi_{\text{b}}\left(i_{1}i_{2},v_{1}v_{3}\right)=\frac{8}{25}\phi_{n}\left(i_{1},v_{1}\right)$.
The largest amount of resource of type $n$ node $i_{1}$ can provision
to $v_{1}$ is then $\phi_{n}\left(i_{1},v_{1}\right)=\frac{25}{26}$,
which leads to $\phi_{\text{b}}(i_{1}i_{2},v_{1}v_{2})=\frac{8}{26}$,
and $\phi_{\text{b}}(i_{1}i_{2},v_{1}v_{3})=\frac{18}{26}$. Considering
$\phi_{n}\left(i_{1},v_{1}\right)=1$ would lead to $\phi_{\text{b}}(i_{1}i_{2},v_{1}v_{2})+\phi_{\text{b}}(i_{1}i_{2},v_{1}v_{3})=\frac{26}{25}>1$,
which is not consistent with \eqref{eq:Prov_C3_Link_Limit}.

In Figure~\ref{fig:Example_Joint}, an SRD graph with a merge topology
is depicted. Through similar calculations, one gets $\phi_{n}\left(i_{1},v_{1}\right)=\frac{1}{2}$,
$\phi_{n}\left(i_{1},v_{2}\right)=\frac{2}{5}$, $\phi_{n}\left(i_{2},v_{3}\right)=1$,
$\phi_{\text{b}}(i_{1}i_{2},v_{1}v_{3})=\frac{9}{15}$, and $\phi_{\text{b}}(i_{1}i_{2},v_{2}v_{3})=\frac{4}{15}$.
The proportions of provisioned infrastructure node and link resources
are then consistent with the proportions of node and link resource
demands. The proportionality of provisioned resources for links entering
or leaving the same vertices is also ensured.

To indicate whether infrastructure nodes have provisioned resources
for some SRD node of slice~$\sigma$, one introduces the sets of
binary variables $\widetilde{\boldsymbol{\Phi}}^{\sigma}=\left\{ \widetilde{\phi}^{\sigma}(i)\right\} _{i\in\mathcal{N}_{\textrm{I}}}$
and $\widetilde{\boldsymbol{\Phi}}=\left\{ \widetilde{\boldsymbol{\Phi}}^{\sigma}\right\} _{\sigma\in\mathcal{S}}$,
which represents node mapping indicators, \textit{i.e.}, $\widetilde{\phi}^{\sigma}(i)=1$
if at least one of the elements of $\{\phi_{\textrm{c}}^{\sigma}(i,v),\phi_{\textrm{s}}^{\sigma}(i,v)\}_{v\in\mathcal{N}_{\textrm{V}}^{\textrm{\ensuremath{\sigma}}}}$
is strictly positive, and $\widetilde{\phi}^{\sigma}(i)=0$ otherwise.
The relation between $\phi_{n}^{\sigma}(i,v)$ and $\widetilde{\phi}^{\sigma}(i)$
is again nonlinear. As in \eqref{eq:Cover_C7_RelationEta}, both quantities
may be linearly related as follows, $\forall i\in\mathcal{N}_{\textrm{I}},\forall\sigma\in\mathcal{S}$,
\begin{equation}
\sum_{v\in\mathcal{N}_{\textrm{V}}^{\textrm{\ensuremath{\sigma}}}}\sum_{n\in\left\{ \textrm{c},\textrm{s}\right\} }\frac{\phi_{n}^{\sigma}\left(i,v\right)}{2\left|\mathcal{N}_{\textrm{V}}^{\textrm{\ensuremath{\sigma}}}\right|}\leq\widetilde{\phi}^{\sigma}(i)<\sum_{v\in\mathcal{N}_{\textrm{V}}^{\textrm{\ensuremath{\sigma}}}}\sum_{n\in\left\{ \textrm{c},\textrm{s}\right\} }\frac{\phi_{n}^{\sigma}\left(i,v\right)}{2\left|\mathcal{N}_{\textrm{V}}^{\textrm{\ensuremath{\sigma}}}\right|}+1.\label{eq:Prov_C8_Node_Ceiling}
\end{equation}

The leasing cost related to the provisioning of computing, storage,
and bandwidth resources in the \emph{wired} part of the infrastructure
network for all slices in $\mathcal{S}$ can be expressed as
\begin{equation}
\begin{split}c_{\textrm{wr}} & \left(\boldsymbol{\Phi}_{\textrm{n}},\widetilde{\boldsymbol{\Phi}},\boldsymbol{\Phi}_{\textrm{b}}\right)=\sum_{\sigma\in\mathcal{S}}c_{\textrm{wr}}^{\sigma}\left(\boldsymbol{\Phi}_{\textrm{n}}^{\sigma},\widetilde{\boldsymbol{\Phi}}^{\sigma},\boldsymbol{\Phi}_{\textrm{b}}^{\sigma}\right)\end{split}
,\label{eq:Prov_Cost}
\end{equation}
with%
\begin{equation}
\begin{split}c_{\textrm{wr}}^{\sigma}\left(\boldsymbol{\Phi}_{\textrm{n}}^{\sigma},\widetilde{\boldsymbol{\Phi}}^{\sigma},\boldsymbol{\Phi}_{\textrm{b}}^{\sigma}\right) & =\sum\limits _{i\in\mathcal{N}_{\textrm{I}}\backslash\mathcal{N}_{\textrm{I}\text{r}}}\widetilde{\phi}{}^{\sigma}(i)c_{\textrm{f}}(i)\\
+ & \sum_{i\in\mathcal{N}_{\textrm{I}}}\sum_{v\in\mathcal{N}_{\textrm{V}}^{\textrm{\ensuremath{\sigma}}}}\sum_{n\in\left\{ \text{c},\text{s}\right\} }a_{n}(i)\phi_{n}^{\sigma}(i,v)c_{n}(i)\\
+ & \sum_{ij\in\mathcal{E}_{\textrm{I}}}\sum_{vw\in\mathcal{E}_{\textrm{V }}^{\textrm{\ensuremath{\sigma}}}}a_{\text{b}}(ij)\phi_{\text{b}}^{\sigma}(ij,vw)c_{\text{b}}(ij),
\end{split}
\label{eq:Prov_Cost_Sigma}
\end{equation}
where the first term represents the cost for deploying VNFs in infrastructure
nodes, while the second and the third term indicate the total cost
for leasing resources from infrastructure nodes and links. In the
first term, the fixed infrastructure node disposal cost related to
RRH nodes is not considered, since it has already been taken into
account in \eqref{eq:Cover_Cost}.

\section{Single-Step vs Two-Step Provisioning\label{sec:Single-step-vs-Two-step}}

The global provisioning problem has to account for storage and computing
constraints, as well as coverage constraints. It leads to the minimization
of the sum of the costs \eqref{eq:Cover_Cost} and \eqref{eq:Prov_Cost}
\begin{equation}
c_{\textrm{tot}}\left(\boldsymbol{\eta},\widetilde{\boldsymbol{\eta}},\boldsymbol{\Phi}_{\textrm{n}},\widetilde{\boldsymbol{\Phi}},\boldsymbol{\Phi}_{\textrm{b}}\right)=c_{\textrm{rr}}\left(\boldsymbol{\eta},\widetilde{\boldsymbol{\eta}}\right)+c_{\textrm{wr}}\left(\boldsymbol{\Phi}_{\textrm{n}},\widetilde{\boldsymbol{\Phi}},\boldsymbol{\Phi}_{\textrm{b}}\right)\label{eq:GlobalCost}
\end{equation}
with the constraints introduced in Sections~\ref{subsec:Formulation_ProvRadio}
and~\ref{subsec:Formulation_ProvOther}. The provisioning algorithm
minimizing \eqref{eq:GlobalCost} and considering all slices jointly
is denoted as $\mathtt{JRN}$ (Joint Radio and Network provisioning).

When the number of variables in ($\boldsymbol{\Phi}_{\textrm{n}},\widetilde{\boldsymbol{\Phi}},\boldsymbol{\Phi}_{\textrm{b}}$)
and ($\boldsymbol{\eta},\widetilde{\boldsymbol{\eta}}$) increases,
the problem may become intractable. Therefore, a two-step provisioning
algorithm, denoted as $\mathtt{CARP}$ (Coverage-Aware Resource Provisioning),
see Algorithm~\ref{algo_1}, is introduced where both terms of \eqref{eq:GlobalCost}
are minimized separately. The \textit{Radio resource Provisionin}g
problem, denoted by $\mathtt{RP}$, involving the radio coverage constrains
introduced in Section~\ref{subsec:Formulation_ProvRadio}, is solved
first. Then, the \textit{Network resource Provisioning}, denoted by
$\mathtt{NP}$, is solved using the solution of the $\mathtt{RP}$
problem and considering the other resource constraints introduced
in Section~\ref{subsec:Formulation_ProvOther}.

When the resource provisioning problem has to be solved for several
slices, each of the $\mathtt{RP}$ and $\mathtt{NP}$ problems can
be addressed either sequentially for each slice, or jointly for all
slices. Let $\mathtt{SR}$ and $\mathtt{JR}$ denote the sequential
and joint $\mathtt{RP}$, and similarly $\mathtt{SN}$ and $\mathtt{JN}$
denote the sequential and joint $\mathtt{NP}$.

\begin{algorithm} \footnotesize
\caption{Coverage-Aware Resource Provisioning\label{algo_1}}

\KwInput{$\mathcal{G}_{\textrm{I}}=(\mathcal{N}_{\textrm{I}},\mathcal{E}_{\textrm{I}}),\mathcal{S},\{\mathcal{G}_{\textrm{V}}^{\sigma},\sigma\in\mathcal{S}\},\{\mathcal{A}^{\sigma},\sigma\in\mathcal{S}\}$}

\KwOutput{$(\widehat{\boldsymbol{\eta}},\widehat{\tilde{\boldsymbol{\eta}}})$
and $(\widehat{\boldsymbol{\Phi}}_{\textrm{n}},\boldsymbol{\widehat{\widetilde{\Phi}}},\widehat{\boldsymbol{\Phi}}_{\textrm{e}})$}

\BlankLine

\textit{\# Radio resource provisioning - JR variant}

Evaluate $(\widehat{\boldsymbol{\eta}},\widehat{\tilde{\boldsymbol{\eta}}})=\arg\min_{\boldsymbol{\eta},\tilde{\boldsymbol{\eta}}}c_{\textrm{rr}}(\boldsymbol{\eta},\tilde{\boldsymbol{\eta}})$,

subject to: \eqref{eq:Cover_C1_EtaLessThanOne},

$\hspace{1cm}$\eqref{eq:Cover_C2_RequiredMinRateUL}-\eqref{eq:Cover_C3_RequiredMinRateDL},
\eqref{eq:Cover_C6_UpDownProportionality}, $\forall\sigma\in\mathcal{S},\forall q\in\mathcal{Q}^{\sigma}$,

$\hspace{1cm}$\eqref{eq:Cover_C4_RadioNodeDemandUL}-\eqref{eq:Cover_C5_RadioNodeDemandDL},
$\forall\sigma\in\mathcal{S}$,

$\hspace{1cm}$\eqref{eq:Cover_C7_RelationEta}, $\forall i\in\mathcal{N}_{\text{Ir}}$,
$\forall\sigma\in\mathcal{S}$.


\textit{\# Radio resource provisioning - SR variant}

\For{$\sigma\in\mathcal{S}$}{

Evaluate $(\widehat{\boldsymbol{\eta}}^{\sigma},\widehat{\tilde{\boldsymbol{\eta}}}^{\sigma})=\arg\min_{\boldsymbol{\eta}^{\sigma},\tilde{\boldsymbol{\eta}}^{\sigma}}c_{\textrm{rr}}^{\sigma}(\boldsymbol{\eta}^{\sigma},\tilde{\boldsymbol{\eta}}^{\sigma})$,

subject to:

$\hspace{1cm}$$\sum\limits _{\sigma'\leqslant\sigma}\sum\limits _{q\in\mathcal{Q}^{\sigma'}}\left(\eta_{\text{u}}^{\sigma'}\left(i,q\right)+\eta_{\text{d}}^{\sigma'}\left(i,q\right)\right)\leqslant1,\,\forall i\in\mathcal{N}_{\text{Ir}}$,

$\hspace{1cm}$\eqref{eq:Cover_C2_RequiredMinRateUL}-\eqref{eq:Cover_C3_RequiredMinRateDL},
\eqref{eq:Cover_C6_UpDownProportionality}, $\forall q\in\mathcal{Q}^{\sigma}$,

$\hspace{1cm}$\eqref{eq:Cover_C4_RadioNodeDemandUL}-\eqref{eq:Cover_C5_RadioNodeDemandDL},

$\hspace{1cm}$\eqref{eq:Cover_C7_RelationEta}, $\forall i\in\mathcal{N}_{\text{Ir}}$.

} 


\textit{\# Other network resource provisioning - JN variant}

Evaluate $(\widehat{\boldsymbol{\Phi}}_{\textrm{n}},\boldsymbol{\widehat{\widetilde{\Phi}}},\widehat{\boldsymbol{\Phi}}_{\textrm{b}})=\arg\min_{\boldsymbol{\Phi}_{\textrm{n}},\widetilde{\boldsymbol{\Phi}},\boldsymbol{\Phi}_{\textrm{b}}}c_{\textrm{wr}}(\boldsymbol{\Phi}_{\textrm{n}},\widetilde{\boldsymbol{\Phi}},\boldsymbol{\Phi}_{\textrm{b}})$

subject to:

$\hspace{1cm}$\eqref{eq:Prov_C1_Node_Satisfied}, $\forall\sigma\in\mathcal{S}$

$\hspace{1cm}$\eqref{eq:Prov_C2_Node_Limit}, \eqref{eq:Prov_C3_Link_Limit},

$\hspace{1cm}$\eqref{eq:Prov_C4_Node_Min_Requirement}-\eqref{eq:Prov_C7_Internal_Link2},
\eqref{eq:Prov_C6_FlowConservation}, $\forall\sigma\in\mathcal{S}$,

$\hspace{1cm}$\eqref{eq:Prov_C8_Node_Ceiling}, $\forall i\in\mathcal{N}_{\textrm{I}}$,
$\forall\sigma\in\mathcal{S}$.

\textit{\# Other network resource provisioning - SN variant}

\For{$\sigma\in\mathcal{S}$}{

Evaluate $(\widehat{\boldsymbol{\Phi}}_{\textrm{n}}^{\sigma},\boldsymbol{\widehat{\widetilde{\Phi}}}^{\sigma},\widehat{\boldsymbol{\Phi}}_{\textrm{b}}^{\sigma})=\arg\min_{\boldsymbol{\Phi}_{\textrm{n}}^{\sigma},\widetilde{\boldsymbol{\Phi}}^{\sigma},\boldsymbol{\Phi}_{\textrm{b}}^{\sigma}}c_{\textrm{wr}}^{\sigma}(\boldsymbol{\Phi}_{\textrm{n}}^{\sigma},\widetilde{\boldsymbol{\Phi}}^{\sigma},\boldsymbol{\Phi}_{\textrm{b}}^{\sigma})$

subject to: \eqref{eq:Prov_C1_Node_Satisfied},

$\hspace{1cm}$$\sum\limits _{\sigma'\leqslant\sigma}\sum\limits _{v\in\mathcal{N}_{\textrm{V}}^{\textrm{\ensuremath{\sigma}}}}\phi_{n}^{\sigma'}(i,v)\leq1,\,\forall n\in\{\textrm{c},\textrm{s}\},\forall i\in\mathcal{N}_{\textrm{I}}$,

$\hspace{1cm}$$\sum\limits _{\sigma'\leqslant\sigma}\sum\limits _{vw\in\mathcal{E}_{\textrm{V }}^{\textrm{\ensuremath{\sigma'}}}}\phi_{\text{b}}^{\sigma'}(ij,vw)\leq1,\,\forall ij\in\mathcal{E}_{\textrm{I}}$,

$\hspace{1cm}$\eqref{eq:Prov_C4_Node_Min_Requirement}-\eqref{eq:Prov_C7_Internal_Link2},
\eqref{eq:Prov_C6_FlowConservation},

$\hspace{1cm}$\eqref{eq:Prov_C8_Node_Ceiling}, $\forall i\in\mathcal{N}_{\textrm{I}}$.

} 

\end{algorithm}

During initialization of $\mathtt{CARP}$, the slice coverage information
$\mathcal{A}^{\sigma}$ is obtained from the SRD, and $\mathcal{A}^{\sigma}$
is partitioned into $Q^{\sigma}$ convex subareas $\mathcal{A}_{q}^{\sigma}$,
$q\in\mathcal{Q}^{\sigma}=\left\{ 1,\dots,Q^{\sigma}\right\} $.

In Step~$1$ (Lines~1-6 (for $\mathtt{JR}$) or Lines~7-14 (for
$\mathtt{SR}$) of Algorithm~\ref{algo_1}), the values of $\boldsymbol{\eta}$
and $\widetilde{\boldsymbol{\eta}}$ minimizing $c_{\textrm{rr}}\left(\boldsymbol{\eta},\widetilde{\boldsymbol{\eta}}\right)$
while satisfying all constraints related to radio provisioning \eqref{eq:Cover_C1_EtaLessThanOne}-\eqref{eq:Cover_C8_EtaUpDown}
are evaluated.

In Step~2 (Line~15-21 (for $\mathtt{JN}$) or Lines~22-29 (for
$\mathtt{SN}$) of Algorithm~\ref{algo_1}), the values of $\boldsymbol{\Phi}_{\textrm{n}},\widetilde{\boldsymbol{\Phi}},\boldsymbol{\Phi}_{\textrm{b}}$
minimizing $c_{\textrm{wr}}\left(\boldsymbol{\Phi}_{\textrm{n}},\widetilde{\boldsymbol{\Phi}},\boldsymbol{\Phi}_{\textrm{b}}\right)$,
subject to the constraints \eqref{eq:Prov_C1_Node_Satisfied}-\eqref{eq:Prov_C8_Node_Ceiling}
are evaluated. The constraints \eqref{eq:Prov_C5_ProportionalityRadio},
\eqref{eq:Prov_RadioDownlink}, \eqref{eq:Prov_RadioUplink} are evaluated
with the help of $\boldsymbol{\eta}$ and $\widetilde{\boldsymbol{\eta}}$
obtained at Step~$1$.

Combining these methods gives four variants of the $\mathtt{CARP}$
provisioning algorithm ($\mathtt{SR}$-$\mathtt{SN}$, $\mathtt{SR}$-$\mathtt{JN}$,
$\mathtt{JR}$-$\mathtt{SN}$, and $\mathtt{JR}$-$\mathtt{JN}$),
as summarized in Table~\ref{tab:Variants} with the number of $\mathtt{RP}$
and $\mathtt{NP}$ problems and the corresponding number of variables
per problem to be handled by each variant. The complexity of the single-step
$\mathtt{JRN}$ algorithm, performing a simultaneous joint radio and
network provisioning for all slices is provided as a reference. In
Table~\ref{tab:Variants}, the variables~$\kappa_{n}^{\sigma}(i,v)$
introduced in \eqref{eq:Prov_C4_Node_Min_Requirement} are not taken
into account, since they are directly related to $\phi_{n}^{\sigma}(i,v)$.

The sequential variants ($\mathtt{SR}$ and $\mathtt{SN}$) require
to solve $|\mathcal{S}|$ optimization problems, but with $|\mathcal{S}|$
less variables compared to the joint variants ($\mathtt{JR}$ and
$\mathtt{JN})$. Since each problem is NP-hard, the sequential variants
may obviously be solved faster than the joint variants. Section~\ref{subsec:SetUp}
compares these variants on simulations.

When the amount of available infrastructure resources is not sufficient
to accommodate all slices, the proposed joint approaches return no
solution. In the sequential approach, the provisioning is performed
slice-by-slice. The first processed requests are likely to be satisfied.
Next requests may only be satisfied when resources are released. This
solution works on a first-arrived-first-served strategy, and has thus
some fairness. The main drawback is the suboptimality of the sequential
approach, which will be discussed in the next section.

Alternatively, in the joint approach, one may renegotiate the SLAs
of all slices to provide some fairness by deploying a part of the
services. This may be done by provisioning resources so as to satisfy
only a fixed proportion $\delta\in]0,1]$ of demands of each slice.
The search for $\delta$ may be done by dichotomy.

\begin{table}[tbh]

\caption{Variants of the Provisioning Algorithm \label{tab:Variants}}

\centering  
\begin{tabular}{    
p{0.1\columnwidth}    
p{0.1\columnwidth}         
p{0.4\columnwidth} }  
\toprule  

\multicolumn{1}{c}{ \textit{Variant} }& \multicolumn{1}{c}{ \#\textit{problems}}& \multicolumn{1}{l}{
\#\textit{variables}/\textit{problem}} \\

\cmidrule[0.4pt](lr{0.12em}){1-1}%
\cmidrule[0.4pt](lr{0.12em}){2-2}%
\cmidrule[0.4pt](lr{0.12em}){3-3}%

\multicolumn{1}{c}{$\mathtt{JRN}$ }& \multicolumn{1}{c}{ $1$ }&
$|\mathcal{S}|(|\mathcal{N}_{\text{Ir}}|\left(1+|\mathcal{Q}^{\sigma}|\right)+$
\\

\multicolumn{1}{c}{}&\multicolumn{1}{c}{}& $2|\mathcal{N}_{\text{I}}||\mathcal{N}_{\text{V}}^{\sigma}|+|\mathcal{N}_{\text{I}}|+|\mathcal{E}_{\text{I}}||\mathcal{E}_{\text{V}}^{\sigma}|)$
\\

\multicolumn{1}{c}{$\mathtt{SR}$-$\mathtt{SN}$ }& \multicolumn{1}{c}{$|\mathcal{S}|$
$\mathtt{RP}$}& $|\mathcal{N}_{\text{Ir}}|\left(1+|\mathcal{Q}^{\sigma}|\right)$
\\

& \multicolumn{1}{c}{$|\mathcal{S}|$ $\mathtt{NP}$ }& $2|\mathcal{N}_{\text{I}}||\mathcal{N}_{\text{V}}^{\sigma}|+|\mathcal{E}_{\text{I}}||\mathcal{E}_{\text{V}}^{\sigma}|$
\\ 

\multicolumn{1}{c}{$\mathtt{SR}$-$\mathtt{JN}$ }&\multicolumn{1}{c}{$|\mathcal{S}|$
$\mathtt{RP}$ }& $|\mathcal{N}_{\text{Ir}}|\left(1+|\mathcal{Q}^{\sigma}|\right)$
\\  

&\multicolumn{1}{c}{$1$ $\mathtt{NP}$ }& $|\mathcal{S}|\left(2|\mathcal{N}_{\text{I}}||\mathcal{N}_{\text{V}}^{\sigma}|+|\mathcal{E}_{\text{I}}||\mathcal{E}_{\text{V}}^{\sigma}|\right)$
\\

\multicolumn{1}{c}{ $\mathtt{JR}$-$\mathtt{SN}$ }&\multicolumn{1}{c}{
$1$ $\mathtt{RP}$ }& $|\mathcal{S}||\mathcal{N}_{\text{Ir}}|\left(1+|\mathcal{Q}^{\sigma}|\right)$
\\

& \multicolumn{1}{c}{ $|\mathcal{S}|$ $\mathtt{NP}$ }& $2|\mathcal{N}_{\text{I}}||\mathcal{N}_{\text{V}}^{\sigma}|+|\mathcal{E}_{\text{I}}||\mathcal{E}_{\text{V}}^{\sigma}|$
\\

\multicolumn{1}{c}{$\mathtt{JR}$-$\mathtt{JN}$ }& \multicolumn{1}{c}{$1$
$\mathtt{RP}$ }& $|\mathcal{S}||\mathcal{N}_{\text{Ir}}|\left(1+|\mathcal{Q}^{\sigma}|\right)$
\\  

&\multicolumn{1}{c}{$1$ $\mathtt{NP}$ }& $|\mathcal{S}|\left(2|\mathcal{N}_{\text{I}}||\mathcal{N}_{\text{V}}^{\sigma}|+|\mathcal{E}_{\text{I}}||\mathcal{E}_{\text{V}}^{\sigma}|\right)$
\\ 

\bottomrule  
\end{tabular}  
\end{table}

\section{Evaluation\label{sec:Evaluation}}

In this section, one evaluates via simulations the performance of
the proposed provisioning algorithms. The simulation set-up is described
in Section~\ref{subsec:SetUp}. The variants of the provisioning
algorithm introduced in Section~\ref{sec:Single-step-vs-Two-step}
are first compared in Section~\ref{subsec:Comparison}. Then, Section~\ref{subsec:Eva_Prov_Benefit}
illustrates the benefits of provisioning prior to SFC embedding compared
to direct SFC embedding. All simulations are performed with the CPLEX
MILP solver interfaced with MATLAB.

\subsection{Simulation Conditions\label{subsec:SetUp}}

\subsubsection{Infrastructure Topology}

Consider the $1.43\,\textrm{km\ensuremath{\times4.95\,\textrm{km}}}$
area around the Stade de France in Seine-Saint-Denis (suburban area
of the city of Paris) shown in Figure~\ref{fig:Orange_Map}. The
map includes real coordinates of RRH nodes (indicated by blue markers)
taken from the open database provided by the French National Agency
of Frequencies\footnote{L'Agence nationale des fr\'equences (ANFR): https://data.anfr.fr/}.

\begin{figure}[tbh]
\begin{centering}
\includegraphics[width=1\columnwidth]{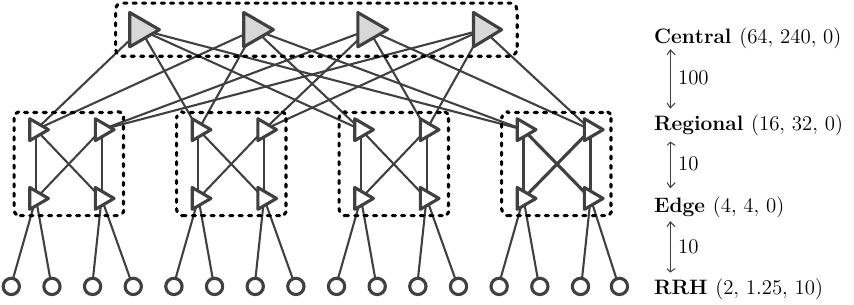}
\par\end{centering}
\caption{Description of the $k$-ary fat-tree infrastructure network in case
$k=4$; Nodes provide a given amount of computing $a_{\textrm{c}}$,
storage $a_{\textrm{s}}$, and radio resource $a_{\text{r}}$ measured
in number of used CPUs, Gbytes, and RBs respectively; Links are assigned
with a given amount of bandwidth $a_{\text{b}}$ measured in Gbps.\label{fig:Description_Infra}}
\end{figure}

For the wired part of the infrastructure network, as in \cite{Riggio2016,Bouten2017},
a $k$-ary fat-tree infrastructure topology is considered, see Figure~\ref{fig:Description_Infra}.
The leaf nodes represent the RRHs. The other nodes represent the edge,
regional, and central data centers. Infrastructure nodes and links
provide a given amount of computing, storage, and possibly radio resources
$\left(a_{\textrm{c}},a_{\textrm{s}},a_{\textrm{r}}\right)$ expressed
in available number of CPUs, Gbytes of storage, and available RBs
at each RRH, depending on the level they belong to.

Only the RRH nodes are represented in Figure~\ref{fig:Orange_Map}.
The locations of the remaining parts of the infrastructure network
(central, regional, and edge nodes) are not displayed. The leasing
costs of each resource of the infrastructure network is detailed in
Table~\ref{tab:Infrastructure-cost}.

\begin{table}[tbh]  
\caption{Infrastructure cost\label{tab:Infrastructure-cost}}
\centering  
\begin{tabular}{    
p{0.1\columnwidth} 
p{0.1\columnwidth}    
p{0.1\columnwidth}         
p{0.1\columnwidth}  
p{0.1\columnwidth}  }  
\toprule  
\multicolumn{1}{c}{\it Node}  
& \multicolumn{1}{c}{$c_{\text{f}}\left(i\right)$} 
& \multicolumn{1}{c}{$c_{\text{r}}\left(i\right)$}
& \multicolumn{1}{c}{$c_{\text{c}}\left(i\right)$} 
& \multicolumn{1}{c}{$c_{\text{s}}\left(i\right)$} \\
\cmidrule[0.4pt](lr{0.12em}){1-1}%
\cmidrule[0.4pt](lr{0.12em}){2-2}%
\cmidrule[0.4pt](lr{0.12em}){3-3}%
\cmidrule[0.4pt](lr{0.12em}){4-4}%
\cmidrule[0.4pt](lr{0.12em}){5-5}%
\multicolumn{1}{c}{$i\in\mathcal{N}_{\text{I}}\backslash\mathcal{N}_{\text{Ir}}$}  
& \multicolumn{1}{c}{$20$} 
& \multicolumn{1}{c}{$-$} 
& \multicolumn{1}{c}{$1$} 
& \multicolumn{1}{c}{$1$} \\ 
\multicolumn{1}{c}{$i\in\mathcal{N}_{\text{Ir}}$} 
& \multicolumn{1}{c}{25} 
& \multicolumn{1}{c}{$0.05$}
& \multicolumn{1}{c}{$1$} 
& \multicolumn{1}{c}{$1$} \\  
\bottomrule  
\end{tabular}  
\end{table}

\subsubsection{Slice Resource Demand (SRD)}

Three types of slices are considered.
\begin{itemize}
\item Slices of type~1 cover the \textit{Stade de France} and aim to provide
an HD video streaming service at $4$~Mbps for at most $200$ VIP
users within the stadium (downlink traffic);
\item Slices of type~2 are dedicated to provide an SD video streaming service
at $0.5$~Mbps, and cover the blue-highlighted area in Figure~\ref{fig:Orange_Map}
(downlink traffic);
\item Slices of type~3 aim to provide a video surveillance and traffic
monitoring service at $1$~Mbps for $50$ cameras installed on the
A86 highway (uplink traffic).
\end{itemize}
The first two slice types address a video streaming service, and thus
have the same function architecture with $3$ virtual functions: a
vVOC, a vGW, and a vBBU. The third slice type consists of five virtual
functions: a vBBU, a vGW, a virtual Traffic Monitor (vTM), a vVOC,
and a virtual Intrusion Detection Prevention System (vIDPS)..

As detailed in Section~\ref{sec:NetworkModel}, the resource requirements
for the various SFCs that will have to be deployed within a slice
are aggregated within an SRD graph that mimics the graph of SFCs.
SRD nodes and links are characterized by the aggregated resource needed
to support the targeted number of users. Details of each resource
type as well as associated SRD graph are given in Table~\ref{tab:SRD}.
Numerical values in Table~\ref{tab:SRD} have been adapted from \cite{Savi2017}.

In the following, different scenarios are considered with an increasing
number of slices whose distribution among each type is given in Table~\ref{tab:Scenarios}.
This represents, \emph{e.g.,} situations where slices of the same
type are provided by different SPs.

\begin{table}[tbh]  
\caption{Number of slices of each type as a function of $|\mathcal{S}|$\label{tab:Scenarios}}
\centering  
\begin{tabular}{    
p{0.1\columnwidth} 
p{0.1\columnwidth}    
p{0.1\columnwidth}         
p{0.1\columnwidth}  
p{0.1\columnwidth}  }  
\toprule  
\multicolumn{1}{c}{$|\mathcal{S}|$}  
& \multicolumn{1}{c}{$4$}  
& \multicolumn{1}{c}{$6$} 
& \multicolumn{1}{c}{$8$} \\

\cmidrule[0.4pt](lr{0.12em}){1-1}%
\cmidrule[0.4pt](lr{0.12em}){2-2}%
\cmidrule[0.4pt](lr{0.12em}){3-3}%
\cmidrule[0.4pt](lr{0.12em}){4-4}%
\multicolumn{1}{c}{Type 1}
& \multicolumn{1}{c}{$2$} 
& \multicolumn{1}{c}{$2$} 
& \multicolumn{1}{c}{$4$}\\ 
\multicolumn{1}{c}{Type 2} 
& \multicolumn{1}{c}{$1$} 
& \multicolumn{1}{c}{$2$} 
& \multicolumn{1}{c}{$1$}\\
\multicolumn{1}{c}{Type 3} 
& \multicolumn{1}{c}{$1$} 
& \multicolumn{1}{c}{$2$}
& \multicolumn{1}{c}{$3$}\\
\bottomrule  
\end{tabular} 
\end{table}

The coverage area $\mathcal{A}^{\sigma}$ associated to each slice
type is partitioned into rectangular subareas $\mathcal{A}_{q}^{\sigma}$
of $90\,\textrm{m\ensuremath{\times103\,\textrm{m}}}$.

Functional structure and resource requirements for the three slice
types are described in Table~\ref{tab:SRD}.

\begin{table}[tbh]  
\scriptsize  
\caption{Slice Resource Demand.} \label{tab:SRD} 
\centering 
\begin{tabular}{  
p{0.05\columnwidth}   
p{0.04\columnwidth}     
p{0.05\columnwidth}  
p{0.04\columnwidth}     
p{0.05\columnwidth}   
p{0.000001\columnwidth}     
p{0.15\columnwidth}  
p{0.04\columnwidth}  }  
\toprule  
\multicolumn{8}{l}{\textbf{Slice 1: HD video streaming}}\\  
\multicolumn{1}{l}{\it Node}    
& \multicolumn{1}{c}{$r_{\textrm{c}}$}   
& \multicolumn{1}{c}{$\underline{r}_{\textrm{c}}$}  
& \multicolumn{1}{c}{$r_{\textrm{s}}$}   
& \multicolumn{1}{c}{$\underline{r}_{\textrm{s}}$}  
&   
& \multicolumn{1}{l}{\it Link}  
& \multicolumn{1}{c}{$r_{\text{b}}$}\\ 

\cmidrule[0.4pt](lr{0.12em}){1-1}%
\cmidrule[0.4pt](lr{0.12em}){2-2}%
\cmidrule[0.4pt](lr{0.12em}){3-3}%
\cmidrule[0.4pt](lr{0.12em}){4-4}%
\cmidrule[0.4pt](lr{0.12em}){5-5}%
\cmidrule[0.4pt](lr{0.12em}){7-7}%
\cmidrule[0.4pt](lr{0.12em}){8-8}%
{vVOC} & {$1.35$} & {$0.14$} & {$3.75$} & {$0.38$} && {vVOC$\rightarrow$vGW} & {$1.0$}\\ 
{vGW} & {$0.23$} & {$0.02$} & {$0.13$} & {$0.01$}  && {vGW$\rightarrow$vBBU} & {$1.0$}\\
{vBBU} & {$1.00$} & {$0.10$} & {$0.13$} & {$0.01$} && & \\

\multicolumn{8}{l}{}\\  
\multicolumn{8}{l}{\textbf{Slice 2: SD video streaming}}\\  
\multicolumn{1}{l}{\it Node}    
& \multicolumn{1}{c}{$r_{\textrm{c}}$}   
& \multicolumn{1}{c}{$\underline{r}_{\textrm{c}}$}   
& \multicolumn{1}{c}{$r_{\textrm{s}}$}   
& \multicolumn{1}{c}{$\underline{r}_{\textrm{s}}$}   
& 
& \multicolumn{1}{l}{\it Link} 
& \multicolumn{1}{c}{$r_{\text{b}}$}\\  
\cmidrule[0.4pt](lr{0.12em}){1-1}%
\cmidrule[0.4pt](lr{0.12em}){2-2}%
\cmidrule[0.4pt](lr{0.12em}){3-3}%
\cmidrule[0.4pt](lr{0.12em}){4-4}%
\cmidrule[0.4pt](lr{0.12em}){5-5}%
\cmidrule[0.4pt](lr{0.12em}){7-7}%
\cmidrule[0.4pt](lr{0.12em}){8-8}%
{vVOC} & {$1.08$} & {$0.11$} & {$1.88$} & {$0.19$} && {vVOC$\rightarrow$vGW} & {$0.5$}\\ 
{vGW} & {$0.18$} & {$0.02$} & {$0.06$} & {$0.01$} && {vGW$\rightarrow$vBBU} & {$0.5$}\\ 
{vBBU} & {$4.00$} & {$0.40$} & {$0.06$} & {$0.01$} &&  & \tabularnewline
\multicolumn{8}{l}{}\\   
\multicolumn{8}{l}{\textbf{Slice 3: Video surveillance and traffic monitoring}}\\   \multicolumn{1}{l}{\it Node}    
& \multicolumn{1}{c}{$r_{\textrm{c}}$}   
& \multicolumn{1}{c}{$\underline{r}_{\textrm{c}}$}   
& \multicolumn{1}{c}{$r_{\textrm{s}}$}  
& \multicolumn{1}{c}{$\underline{r}_{\textrm{s}}$}  
& 
& \multicolumn{1}{l}{\it Link}  
& \multicolumn{1}{c}{$r_{\text{b}}$}\\ 
\cmidrule[0.4pt](lr{0.12em}){1-1}%
\cmidrule[0.4pt](lr{0.12em}){2-2}%
\cmidrule[0.4pt](lr{0.12em}){3-3}%
\cmidrule[0.4pt](lr{0.12em}){4-4}%
\cmidrule[0.4pt](lr{0.12em}){5-5}%
\cmidrule[0.4pt](lr{0.12em}){7-7}%
\cmidrule[0.4pt](lr{0.12em}){8-8}%
 
{vIDPS} & {$0.535$} & {$0.054$} & {$0.006$} & {$0.001$} && {vIDPS$\rightarrow$vVOC} & {$0.05$}\\
{vVOC} & {$0.270$} & {$0.027$} & {$0.188$} & {$0.019$}  && {vVOC$\rightarrow$vTM} & {$0.05$}\\
{vTM} & {$0.665$} & {$0.067$} & {$0.006$} & {$0.001$}   && {vTM$\rightarrow$vGW} & {$0.05$}\\ 
{vGW} & {$0.045$} & {$0.005$} & {$0.006$} & {$0.001$}   && {vGW$\rightarrow$vBBU} & {$0.05$}\\
{vBBU} & {$0.200$} & {$0.020$} & {$0.006$} & {$0.001$}  &&  & \\
\bottomrule  
\end{tabular}  
\end{table}

\subsubsection{Rate Function\label{subsec:Rate-Allocation-Model}}

The model of $b_{\textrm{d}}\left(x_{i}^{\textrm{r}},\mathcal{A}_{q}^{\sigma}\right)$
and $b_{\textrm{u}}\left(x_{i}^{\textrm{r}},\mathcal{A}_{q}^{\sigma}\right)$
(mentioned in Section~\ref{subsec:Formulation_ProvRadio}), for the
amount of data carried by an RB for a user located in $\mathcal{A}_{q}^{\sigma}$
and served by an RRH located in $x_{i}^{\textrm{r}}$, are now considered.

Let $d\left(x_{i}^{\textrm{r}},\mathcal{A}_{q}^{\sigma}\right)$ be
the distance between $x_{i}^{\textrm{r}}$ and the center of each
rectangle $\mathcal{A}_{q}^{\sigma}$. Focusing on downlink traffic,
according to \cite{Tse2004}, one assumes that
\begin{equation}
b_{\textrm{d}}\left(x_{i}^{\textrm{r}},\mathcal{A}_{q}^{\sigma}\right)=W_{i}\log_{2}\left(1+\cfrac{P_{\text{rx,d}}\left(d\left(x_{i}^{\textrm{r}},\mathcal{A}_{q}^{\sigma}\right)\right)}{P_{\text{n}}}\right),\label{eq:Bd}
\end{equation}
where $W_{i}$ is the bandwidth (in Hz) of an RB provided by RRH~$i$,
$P_{\text{n}}$ is the noise power given by $P_{\text{n}}=W_{i}N_{0}$,
where $N_{0}$ is the noise power spectral density. $P_{\text{rx}}(d)$
is the obtained signal power at the receiver evaluated as 
\begin{equation}
P_{\text{rx,d}}(d)=P_{\text{tx,d}}+G_{\text{tx,d}}+G_{\text{rx,d}}-PL(d),
\end{equation}
where $P_{\text{tx}}$ is the transmission power of the transmitter,
$G_{\text{tx}}$ and $G_{\text{rx}}$ are the antenna gains of the
transmitter and the receiver, and $PL(d)$ is the Path Loss given
by the adapted $\alpha\beta\gamma$-model introduced in \cite{Sun2016a}
for 5G mobile network
\begin{equation}
PL(d)=10\alpha\log_{10}(d)+\beta+10\gamma\log_{10}(f_{i}),
\end{equation}
where $\alpha$ and $\gamma$ are respectively coefficients accounting
for the dependency of the path loss with distance and frequency $f_{i}$,
$\beta$ is an optimized offset value for path loss (dB). $PL$, $d$,
and $f_{i}$ are expressed in dB, meters, and GHz, respectively. An
expression similar to \eqref{eq:Bd} may be derived for $b_{\textrm{u}}\left(x_{i}^{\textrm{r}},\mathcal{A}_{q}^{\sigma}\right)$.

All RRH~$i\in\mathcal{N}_{\textrm{Ir}}$ and all UEs are assumed
to be identical. The parameters for the models $b_{\textrm{d}}\left(x_{i}^{\textrm{r}},\mathcal{A}_{q}^{\sigma}\right)$
and $b_{\textrm{u}}\left(x_{i}^{\textrm{r}},\mathcal{A}_{q}^{\sigma}\right)$
are summarized in Table~\ref{tab:RRH_parameters} and have been partly
taken from \cite{ETSI2016}.\begin{table}[tbh] 
\caption{Parameters of RRH, UE, and $\alpha\beta\gamma$-model.\label{tab:RRH_parameters}}
\centering 
\begin{tabular}{   
p{0.1\columnwidth}   
p{0.5\columnwidth}        
p{0.2\columnwidth} } 
\toprule 
\multicolumn{1}{c}{\it Parameter}   
& \multicolumn{1}{l}{\it Definition}  
& \multicolumn{1}{l}{\it Value} \\ 
\cmidrule[0.4pt](lr{0.12em}){1-1}%
\cmidrule[0.4pt](lr{0.12em}){2-2}%
\cmidrule[0.4pt](lr{0.12em}){3-3}%
\multicolumn{1}{c}{$a_{\textrm{r}}(i)$} & Number of RBs available at RRH~$i$  
& \multicolumn{1}{l}{$100$} \\ 
\multicolumn{1}{c}{$f_{i}$} 
     & Carrier frequency of RRH~$i$       & \multicolumn{1}{l}{$2.6$ GHz} \\ 
\multicolumn{1}{c}{$W_{i}$} 
     & Bandwidth of a RB of RRH~$i$       & \multicolumn{1}{l}{$0.2$ MHz} \\ 
\multicolumn{1}{c}{$P_{\text{tx,d}}$} 
     & Antenna transmit power of each RRH &\multicolumn{1}{l}{ $43$ dBm} \\ 
\multicolumn{1}{c}{$G_{\text{tx,d}}$} 
     & Antenna gain of each RRH          & \multicolumn{1}{l}{$15$ dBi} \\ 
\multicolumn{1}{c}{$P_{\text{tx,u}}$} 
     & Antenna transmit power of each UE & \multicolumn{1}{l}{$23$ dBm} \\ 
\multicolumn{1}{c}{$G_{\text{tx,u}}$} 
     & Antenna gain of each UE           & \multicolumn{1}{l}{$3$ dBi} \\ 
\multicolumn{1}{c}{$N_{0}$} 
     & Noise power spectral density     & \multicolumn{1}{l}{$-174$ dBm/Hz} \\ 
\multicolumn{1}{c}{$\left(\alpha,\beta,\gamma\right)$} 
     & $\alpha\beta\gamma$-model parameters & \multicolumn{1}{l}{$\left(3.6,7.6,2\right)$} \\ 
\bottomrule 
\end{tabular} 
\end{table} 

\subsection{Comparison of Provisioning Algorithms\label{subsec:Comparison}}

This section illustrates the performance of the $\mathtt{JRN}$ joint
approach and of the four variants of the $\mathtt{CARP}$ two-step
provisioning algorithm described in Table~\ref{tab:Variants} when
four, six, and eight slices of different types have to be deployed,
see Table~\ref{tab:Scenarios}.

\begin{figure}[H]
\begin{centering}
\subfloat[\label{fig:NewRes_CoverCost}]{\begin{centering}
\includegraphics[width=0.23\textwidth]{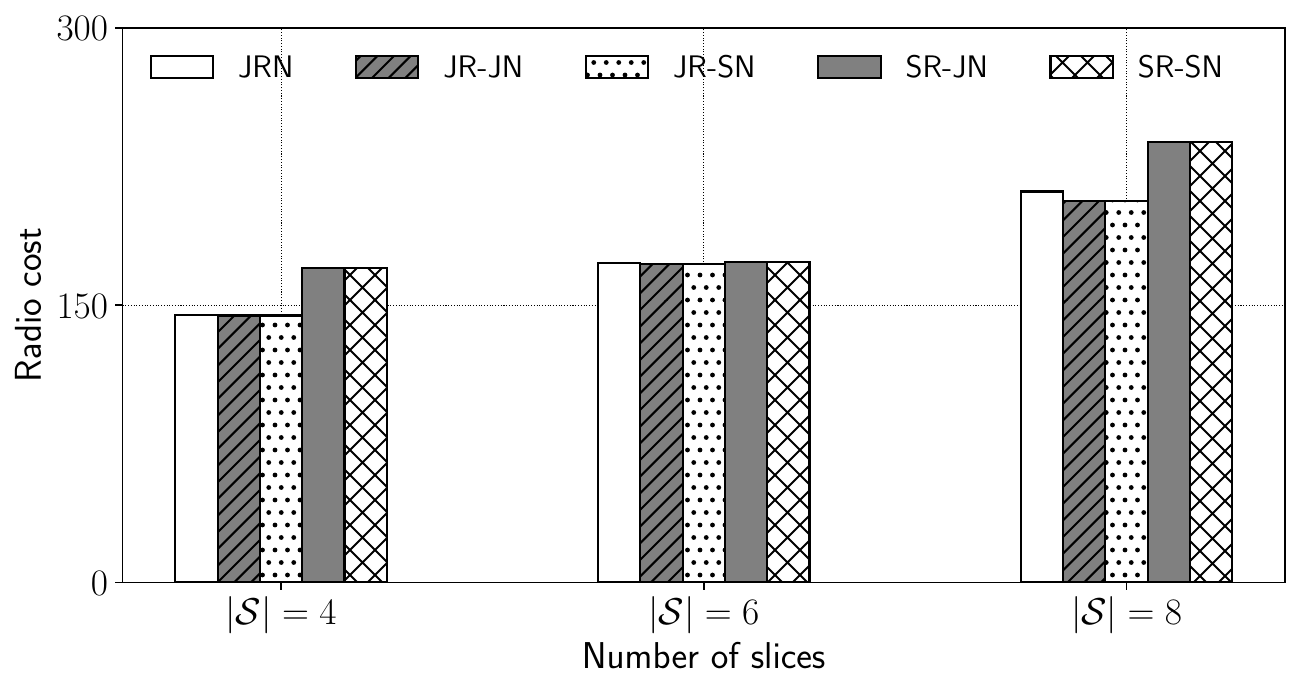}
\par\end{centering}
}\subfloat[\label{fig:NewRes_ProvCost}]{\begin{centering}
\includegraphics[width=0.23\textwidth]{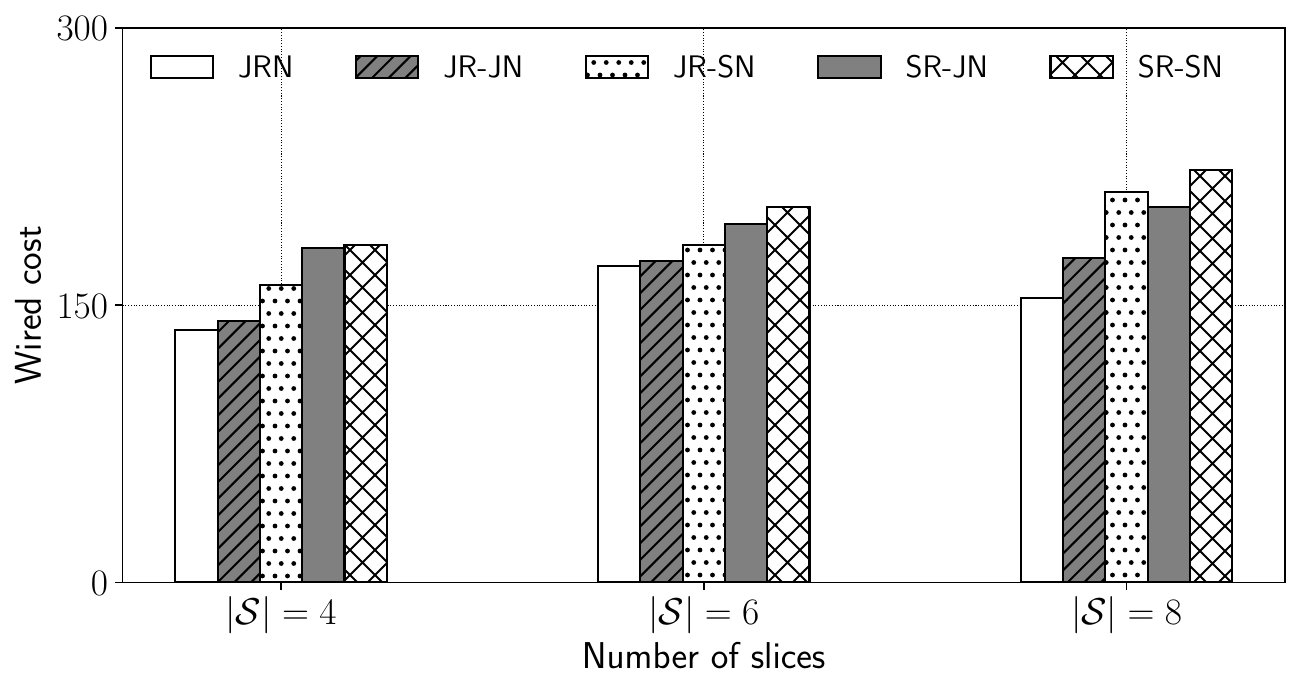}
\par\end{centering}
}
\par\end{centering}
\begin{centering}
\subfloat[\label{fig:NewRes_TotalCost}]{\begin{centering}
\includegraphics[width=0.23\textwidth]{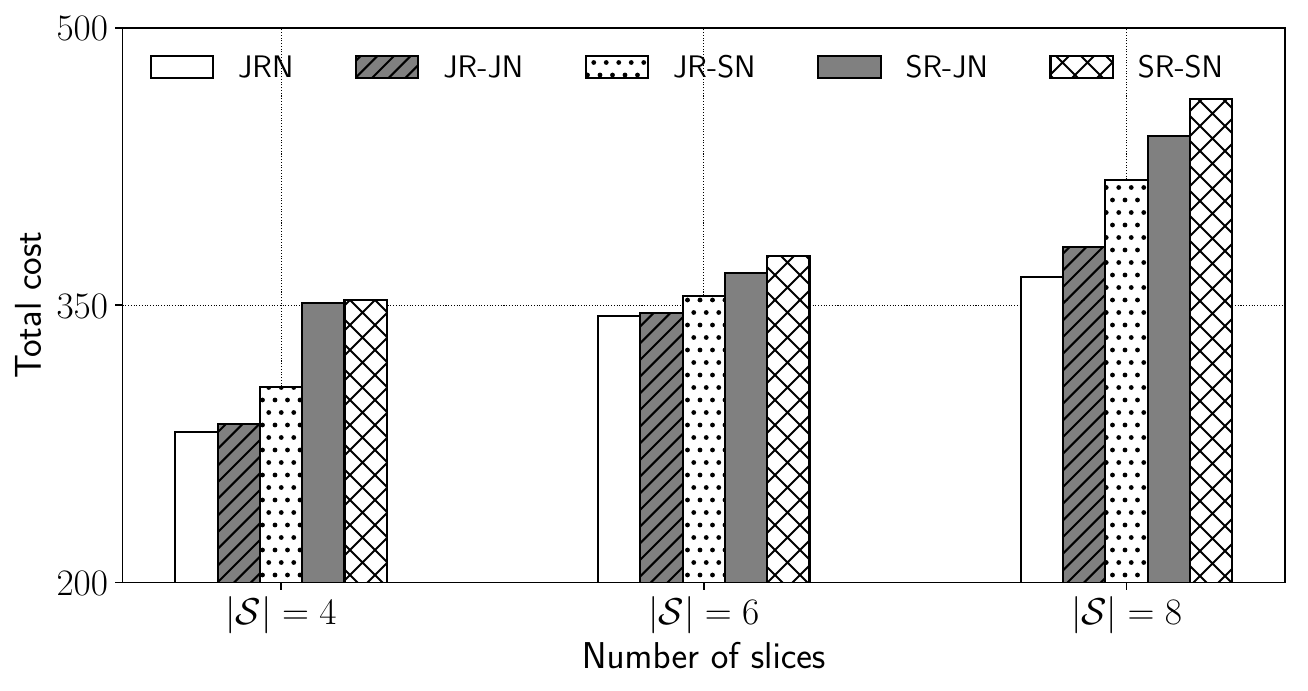}
\par\end{centering}
}\subfloat[\label{fig:NewRes_PerBlock}]{\begin{centering}
\includegraphics[width=0.23\textwidth]{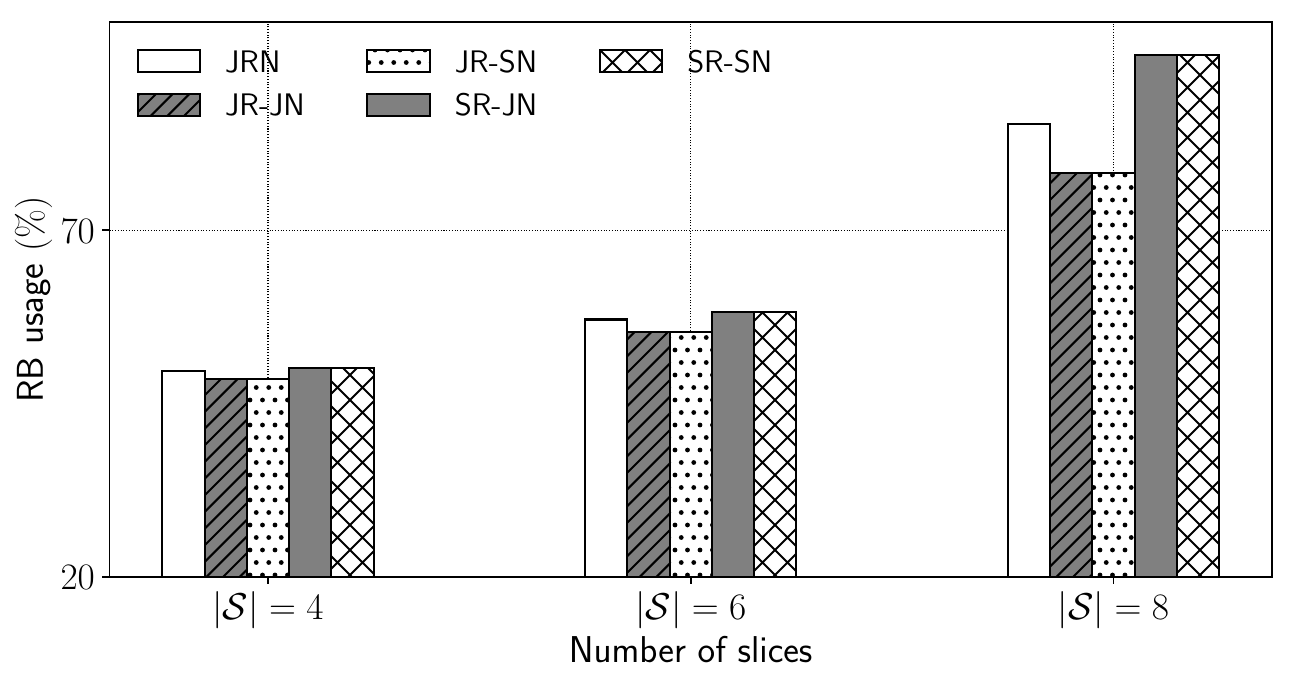}
\par\end{centering}
}
\par\end{centering}
\begin{centering}
\subfloat[\label{fig:NewRes_PerNode}]{\begin{centering}
\includegraphics[width=0.23\textwidth]{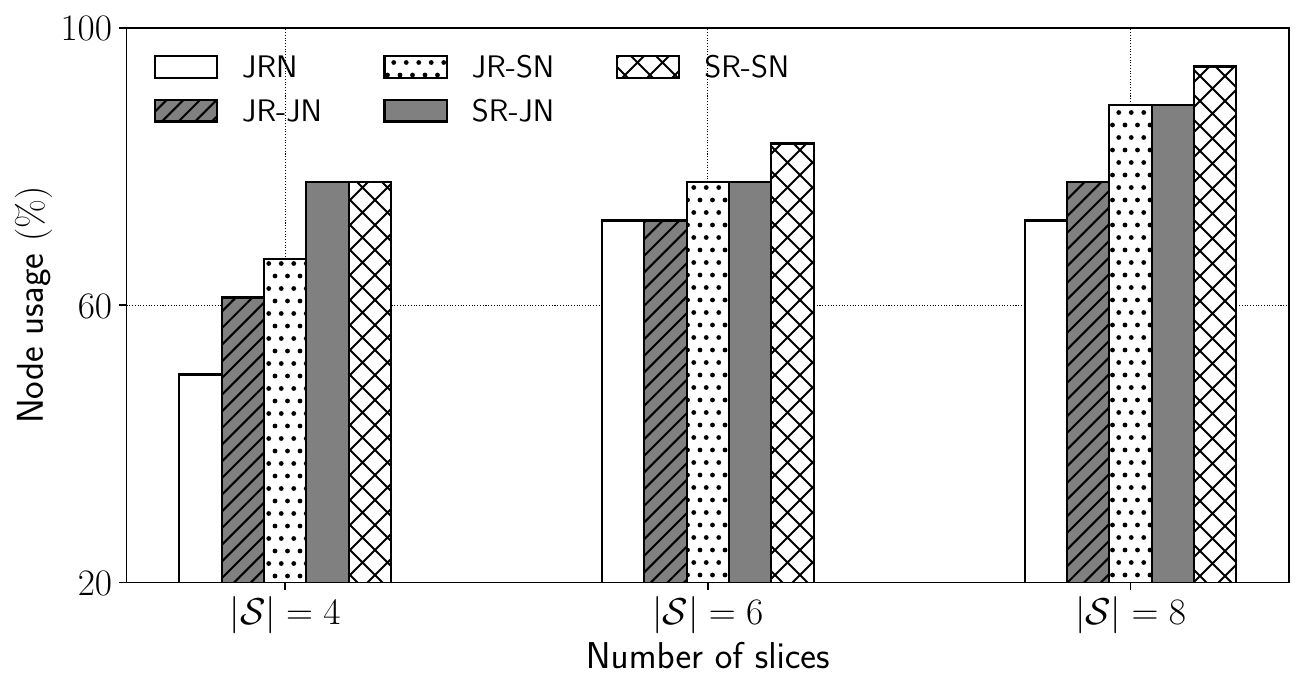}
\par\end{centering}
}\subfloat[\label{fig:NewRes_PerLink}]{\begin{centering}
\includegraphics[width=0.23\textwidth]{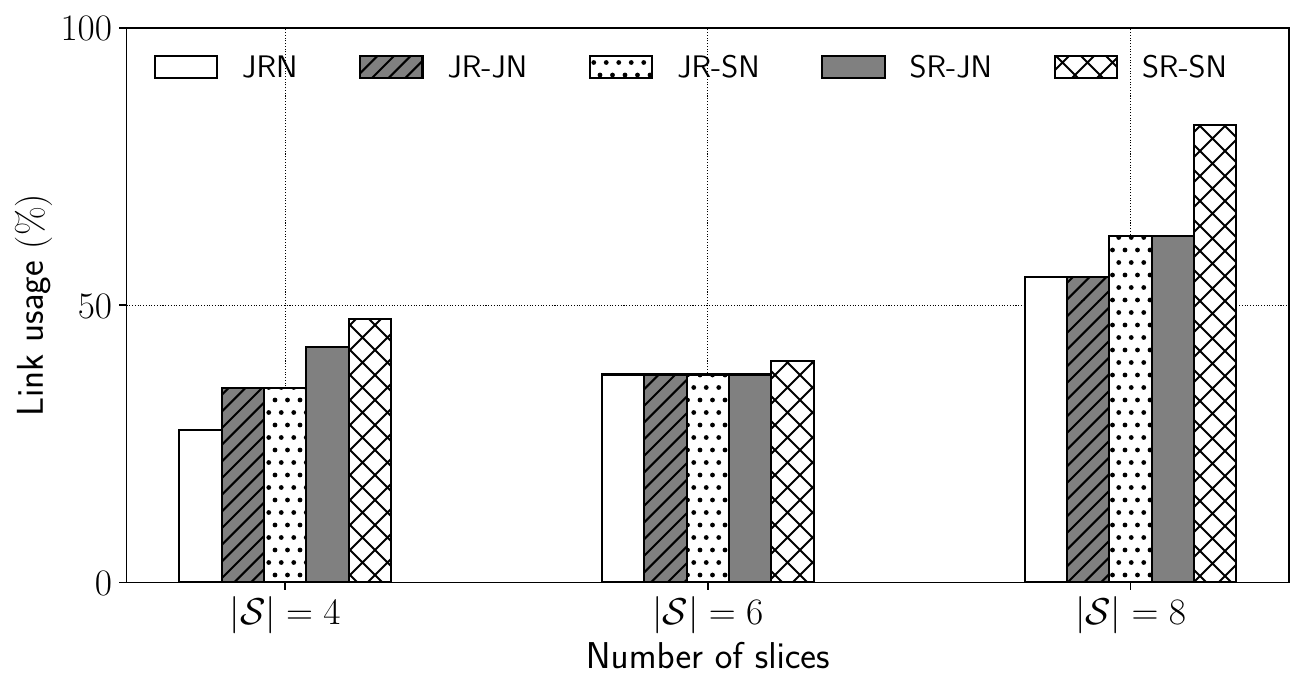}
\par\end{centering}
}
\par\end{centering}
\caption{Performance comparison of $4$ variants in terms of (a) radio cost,
(b) wired cost, (c) total provisioning cost, utilization of (d) RBs,
(e) infrastructure nodes, and (f) infrastructure links. \label{fig:NewRes_Compare_4_Variants}}
\end{figure}

Figure~\ref{fig:NewRes_CoverCost} illustrates the radio provisioning
costs obtained with the various approaches. One observes that the
joint $\mathtt{RP}$ schemes ($\mathtt{JRN}$, $\mathtt{JR}$-$\mathtt{SN}$,
and $\mathtt{JR}$-$\mathtt{JN}$) yield a smaller cost whatever the
$\mathtt{NP}$ allocation method. Note that the $\mathtt{JRN}$ scheme
provides a wireless provisioning cost slightly larger than that of
the $\mathtt{JR}$-$\mathtt{SN}$ or $\mathtt{JR}$-$\mathtt{JN}$
approaches.

Figure~\ref{fig:NewRes_ProvCost} illustrates the cost related to
the wired part of the infrastructure network. The $\mathtt{JRN}$
scheme provides the best results and is always able to compensate
for the somewhat larger radio provisioning cost, as illustrated in
Figure~\ref{fig:NewRes_TotalCost}, which shows the total provisioning
costs. Considering the suboptimal approaches, Figures~\ref{fig:NewRes_ProvCost}
and~\ref{fig:NewRes_TotalCost} show that the $\mathtt{JR}$-$\mathtt{JN}$
scheme performs better than the other approaches and $\mathtt{SR}$-$\mathtt{SN}$
provides always the largest costs, as expected.

To explain these results, one may consider first the use of radio
resource blocks detailed in Figure~\ref{fig:NewRes_PerBlock}. The
results are consistent with those in Figure~\ref{fig:NewRes_CoverCost}:
the joint $\mathtt{RP}$ approaches ($\mathtt{JRN}$, $\mathtt{JR}$-$\mathtt{SN}$,
and $\mathtt{JR}$-$\mathtt{JN}$) outperform the sequential approaches
($\mathtt{SR}$-$\mathtt{JN}$ and $\mathtt{SR}$-$\mathtt{SN}$),
since the joint $\mathtt{RP}$ aims at finding the optimal wireless
provisioning for all the slices, while the sequential method only
accounts for the constraints of each slice sequentially. The $\mathtt{JRN}$
approach does not select the \textit{\emph{best}} RRHs for the radio
resource provisioning, as compared to the $\mathtt{JR}$-$\mathtt{JN}$
or the $\mathtt{JR}$-$\mathtt{SN}$ approach, but rather selects
the\textit{ }RRHs so as to facilitate the wired network resource provisioning.
This leads to a slightly higher utilization of RBs and radio cost
(see Figures~\ref{fig:NewRes_PerBlock} and \ref{fig:NewRes_CoverCost}),
but lower utilization of infrastructure nodes and links (see Figures~\ref{fig:NewRes_PerNode}
and \ref{fig:NewRes_PerLink}), and finally allows the $\mathtt{JRN}$
approach to obtain the lowest total cost.

For the suboptimal approaches, the joint $\mathtt{RP}$ approach also
leads to an efficient utilization of infrastructure nodes and links
when solving the $\mathtt{NP}$ problem, as shown in Figures~\ref{fig:NewRes_PerNode}
and \ref{fig:NewRes_PerLink}.

The difference in performance of these two sets of methods ($\mathtt{JR}$-$\mathtt{SN}$
and $\mathtt{JR}$-$\mathtt{JN}$ versus $\mathtt{SR}$-$\mathtt{JN}$
and $\mathtt{SR}$-$\mathtt{SN}$) becomes more significant when the
number of slices increases. For instance, with six slices, a difference
of $11.11\%$ in terms of link utilization is observed in favor of
the $\mathtt{JR}$-$\mathtt{JN}$ approach, see Figure~\ref{fig:NewRes_PerLink},
whereas with eight slices, the difference is $16.67\%$. Overall,
the $\mathtt{JR}$-$\mathtt{JN}$ approach provides the best performance
in terms of provisioning costs among the four suboptimal methods.

\begin{figure}[tbh]
\begin{centering}
\includegraphics[width=0.5\columnwidth]{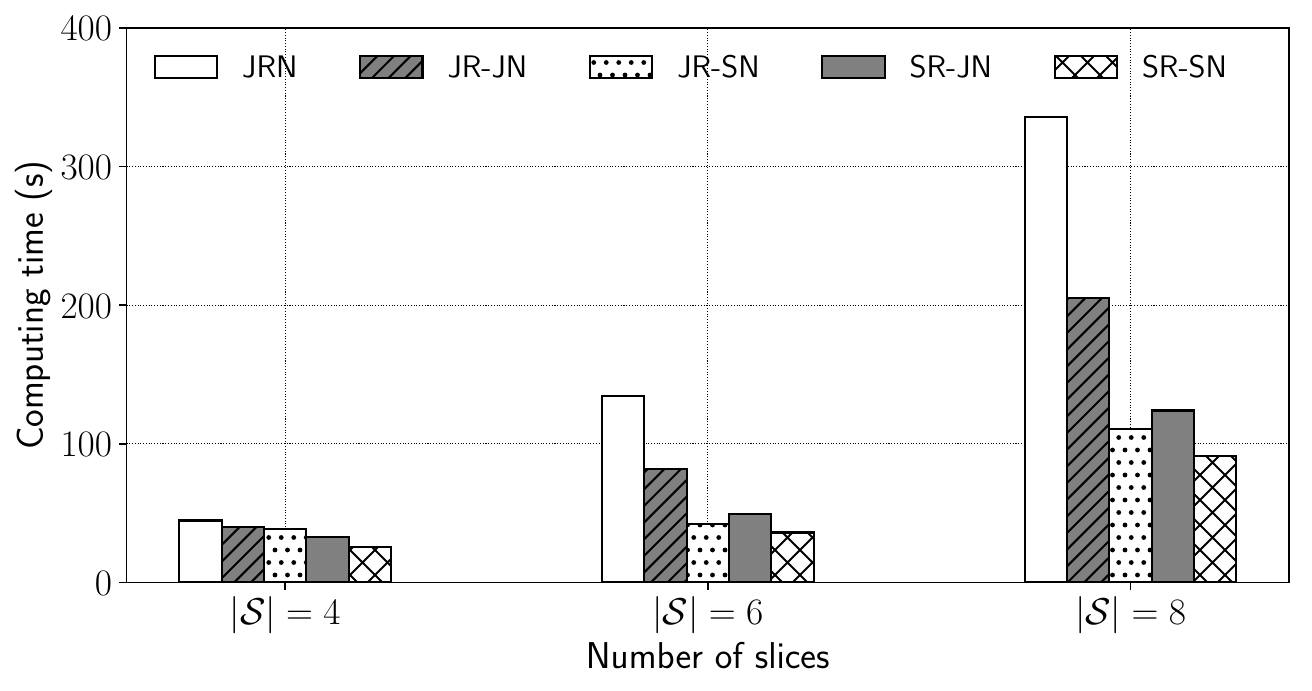}
\par\end{centering}
\caption{Computing time of the $4$ proposed provisioning variants\label{fig:NewRes_Time}}
\end{figure}

As expected, the methods involving sequential provisioning ($\mathtt{SR}$
and $\mathtt{SN}$) perform better than the joint approaches ($\mathtt{JR}$,
$\mathtt{JN}$, and $\mathtt{JRN}$) in terms of computing time. Increasing
the number of slices leads to an increase of the cardinality of the
sets of variables $\left(\boldsymbol{\eta},\tilde{\boldsymbol{\eta}}\right)$
and $\left(\boldsymbol{\Phi}_{\textrm{n}},\widetilde{\boldsymbol{\Phi}},\boldsymbol{\Phi}_{\textrm{b}}\right)$
and therefore increases the computing time. In sequential provisioning,
slices are considered successively. Therefore, among the four suboptimal
methods, the $\mathtt{SR}$-$\mathtt{SN}$ approach and the $\mathtt{JR}$-$\mathtt{JN}$
approach are respectively the least and most time-consuming, as shown
in Figure~\ref{fig:NewRes_Time}. The computing time of the optimal
$\mathtt{JRN}$ is up to $4$ times larger than that of the $\mathtt{SR}$-$\mathtt{SN}$
approach. Moreover, it increases faster than the other approaches
when the number of slices increases.

\begin{figure}[tbh]
\begin{centering}
\includegraphics[width=0.8\columnwidth]{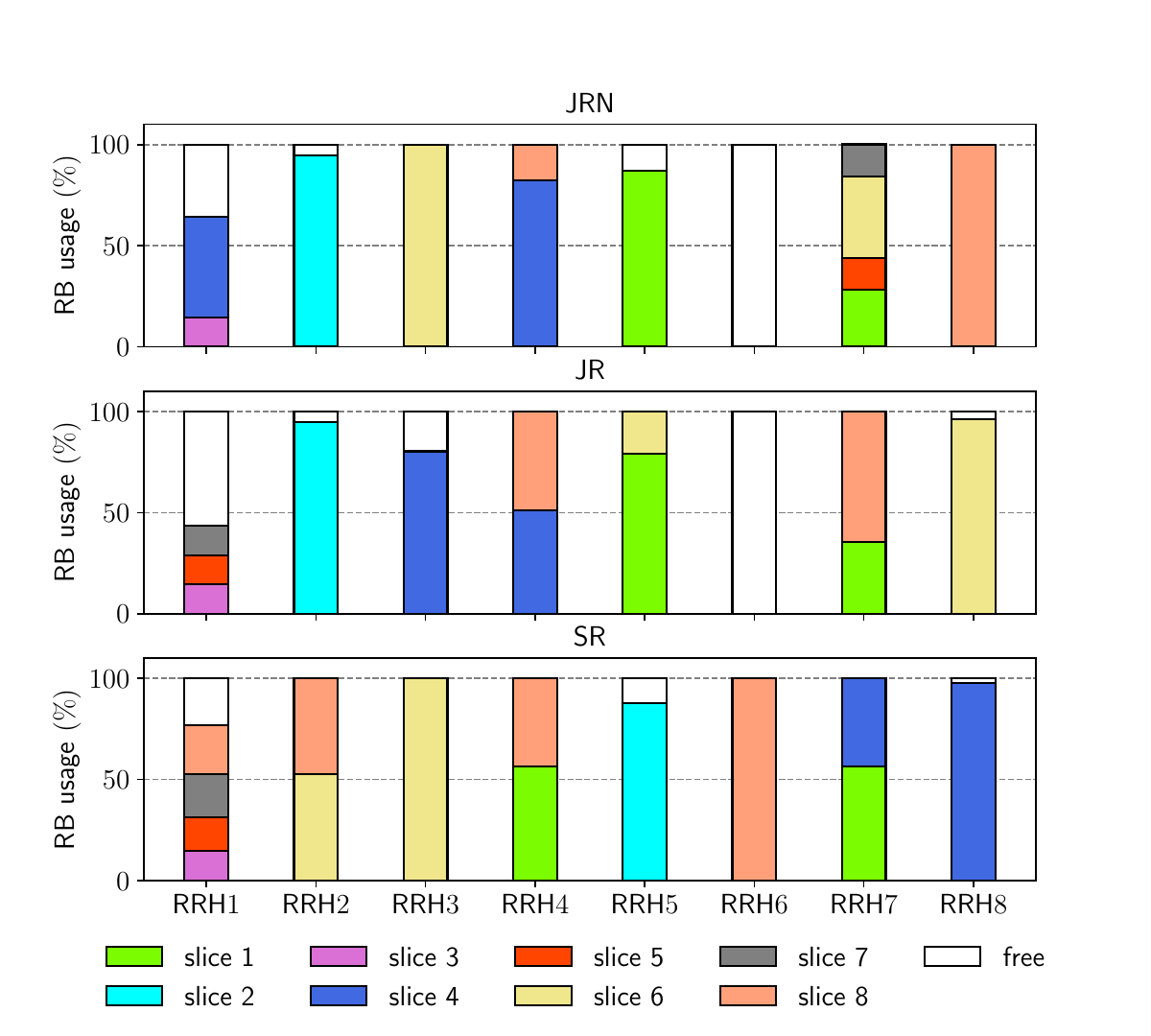}
\par\end{centering}
\caption{Provisioned RBs by RRHs for each slice considering the $\mathtt{JRN}$
(top), the $\mathtt{JR}$ (middle), and the $\mathtt{SR}$ (bottom)
approaches.\label{fig:k_PerBlockDetailed}}
\end{figure}

Figure~\ref{fig:k_PerBlockDetailed} illustrates the way RBs are
provisioned by the various RRHs for each slice, when $|\mathcal{N}_{\text{Ir}}|=8$
and $|\mathcal{S}|=8$. Thanks to the rate-related discount introduced
in the objective function, RRHs that are close to the coverage area
of each slice are chosen in priority. For instance, with the $\mathtt{JRN}$
and $\mathtt{JR}$ approach, Slice 1, which covers the stadium, has
its radio resource demand provisioned by RRH 5 and RRH 7. With the
$\mathtt{SR}$ approach, radio resource demand of Slice 1 is provisioned
by RRH 4 and RRH 7. These three RRHs are both close to the stadium.%

The advantage of the $\mathtt{JRN}$ and $\mathtt{JR}$ over the $\mathtt{SR}$
approach can be observed: with the $\mathtt{SR}$ approach, all RRHs
are required to provision resources, whereas with the $\mathtt{JRN}$
or $\mathtt{JR}$ approach, only seven RRHs are needed.

\begin{figure}[tbh]
\begin{centering}
\includegraphics[width=0.5\columnwidth]{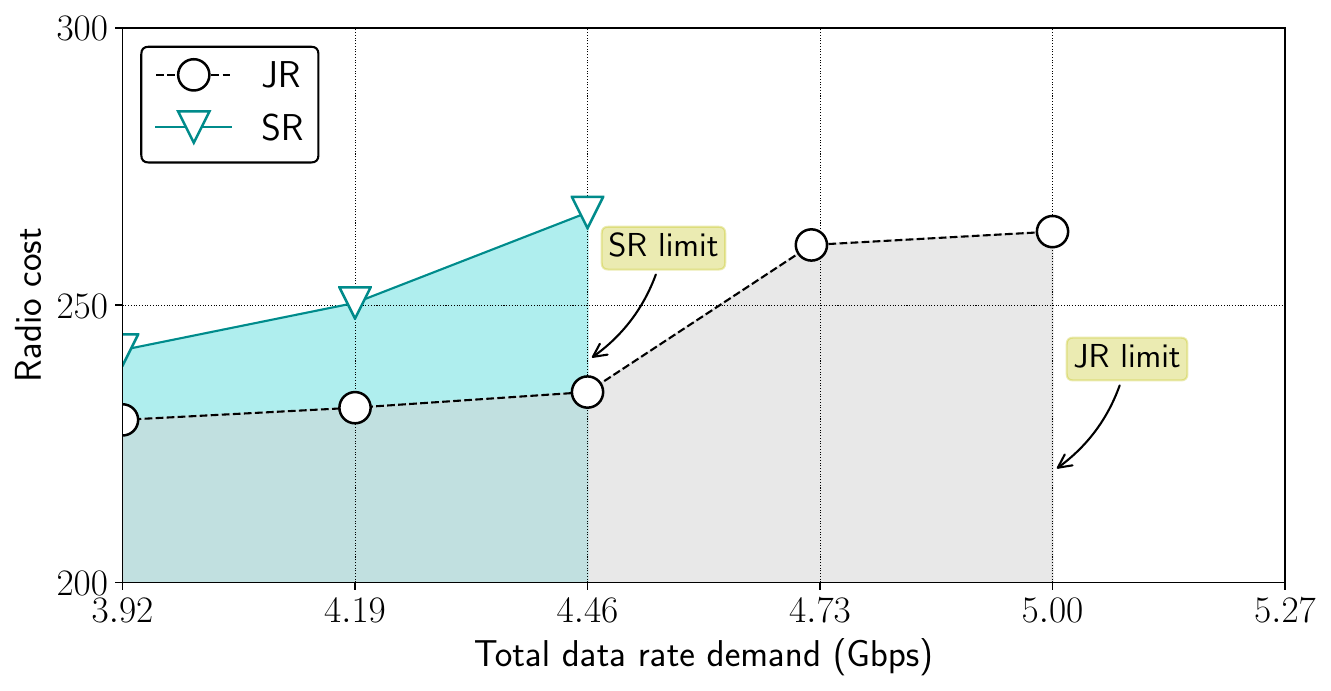}
\par\end{centering}
\caption{Maximum supported data rate associated to the $\mathtt{SR}$ and $\mathtt{JR}$
provisioning approaches when 3 slices of type~1, 2, and 3 have to
be deployed.\label{fig:NewRes_CoverCost_Limit}}
\end{figure}

Finally, Figure~\ref{fig:NewRes_CoverCost_Limit} focuses on the
$\mathtt{RP}$ problem and shows the maximum supported data rate in
the case of sequential and joint radio resource provisioning (\textit{i.e.,}
$\mathtt{SR}$ and $\mathtt{JR}$) as a function of the aggregated
data rate demand from users, \textit{i.e.}, $\sum_{\sigma\in\mathcal{S}}u^{\sigma}\underline{R}^{\sigma}$,
where $u^{\sigma}$ is the number of users in $\sigma$, when $\left|\mathcal{N}_{\text{Ir}}\right|=8$
and 3 slices of type~1, 2, and 3 have to be deployed. $\underline{R}^{\sigma}$
remains constant for each slice $\sigma$. The total number of users
$u^{\sigma}$ associated to each slice varies, but their relative
proportions among slices remain constant. With the $\mathtt{JR}$
approach, a larger aggregated data rate is supported: provisioning
of slices with more users is then possible.

\subsection{Resource Provisioning vs Direct Embedding \label{subsec:Eva_Prov_Benefit}}

In this section, one assumes that the radio provisioning step has
been performed and one focuses on the wired part of the provisioning
problem.

To evaluate the benefits of a provisioning approach prior to SFC embedding,
the latter is compared to a direct SFC embedding approach. A single
slice of type~1 is considered.

For the SFC deployment, the ILP-based SFC embedding algorithm is adapted
from \cite{Riggio2016}. Specifically, the objective function in \cite{Riggio2016}
is modified to allow the simultaneous embedding of multiple SFCs.
Both sequential and joint SFC embedding schemes are performed. The
proposed methods, where provisioning is done before a joint and sequential
SFC embedding, are denoted respectively as $\mathtt{prov}$-$\mathtt{joint}$-$\mathtt{emb}$
and $\mathtt{prov}$-$\mathtt{seq}$-$\mathtt{emb}$. Direct joint
and sequential SFC embedding are denoted as $\mathtt{dir}$-$\mathtt{joint}$-$\mathtt{emb}$
and $\mathtt{dir}$-$\mathtt{seq}$-$\mathtt{emb}$, where prior provisioning
is not considered.

The $k$-ary fat-tree infrastructure topology considered in Section~\ref{subsec:SetUp}
is used here again. The amount of network infrastructure resource
available at each node and link of the infrastructure remains the
same.

\begin{figure}[tbh]
\begin{centering}
\subfloat[\label{fig:RES_Compare_Cost}]{\begin{centering}
\includegraphics[width=0.47\columnwidth]{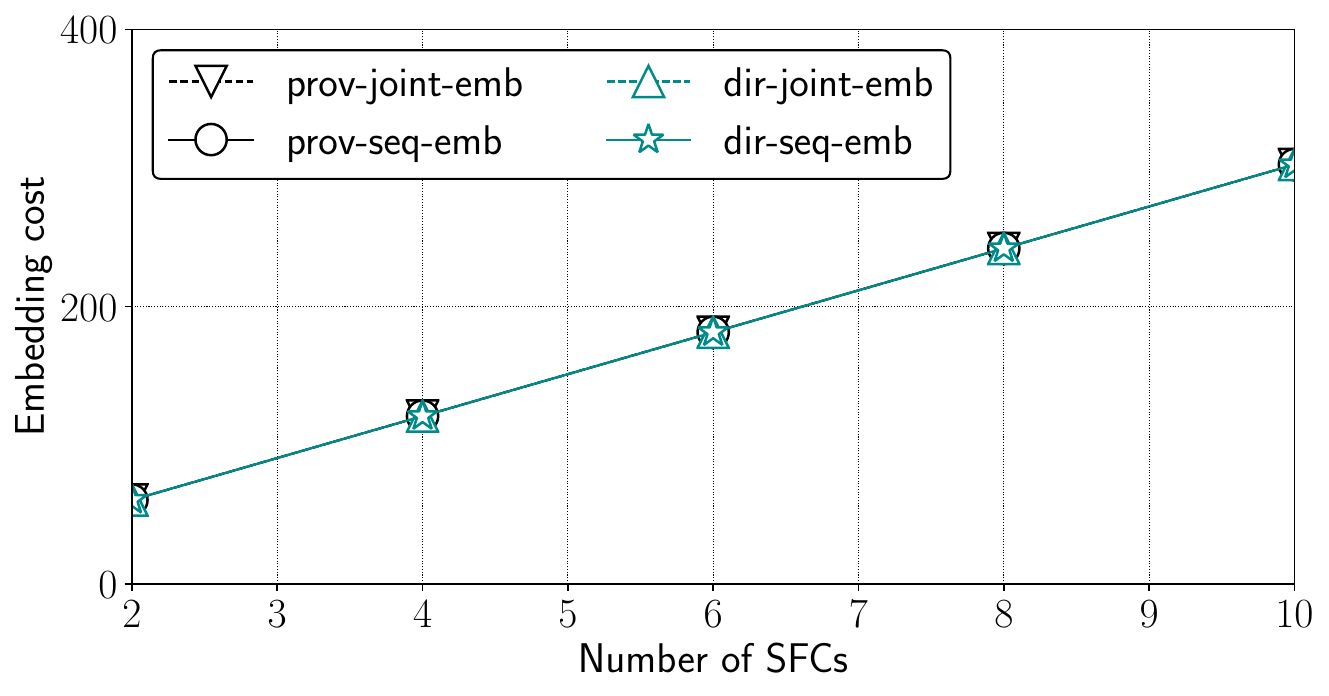}
\par\end{centering}
}\subfloat[\label{fig:RES_Compare_Time}]{\begin{centering}
\includegraphics[width=0.47\columnwidth]{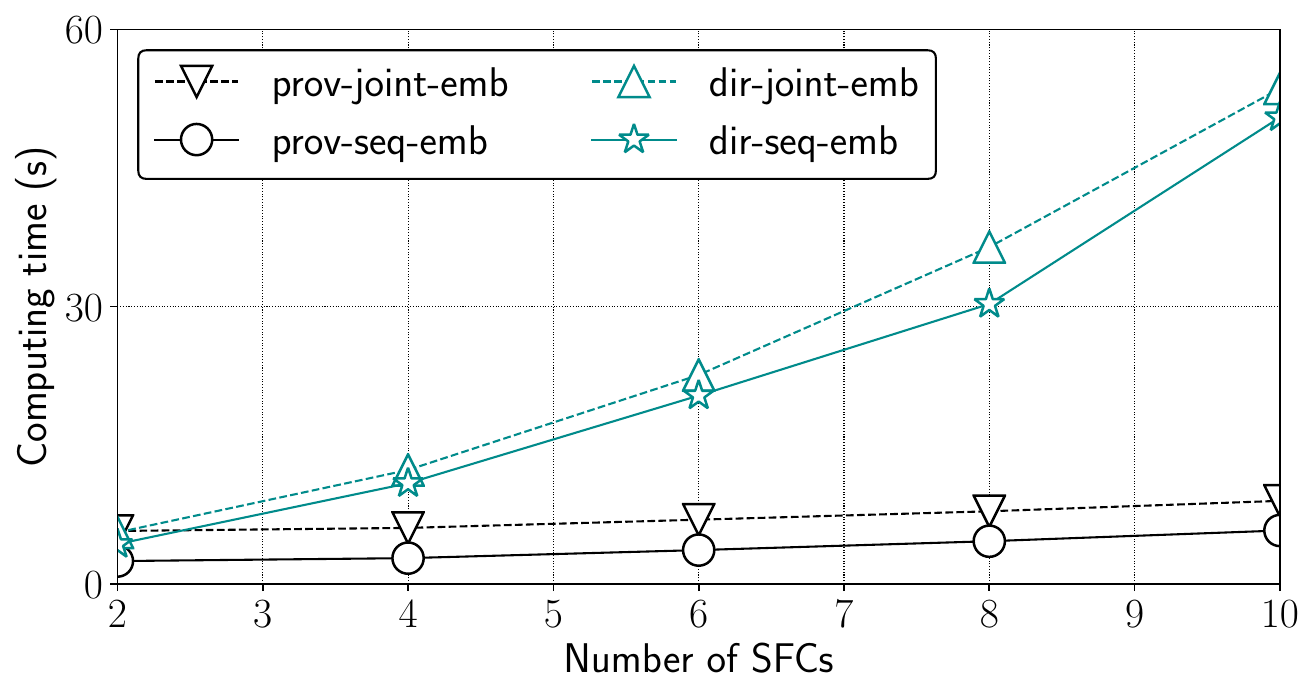}
\par\end{centering}
}
\par\end{centering}
\caption{(a) Embedding costs and (b) computing time of $\mathtt{prov}$-$\mathtt{joint}$-$\mathtt{emb}$,
$\mathtt{prov}$-$\mathtt{seq}$-$\mathtt{emb}$, $\mathtt{dir}$-$\mathtt{joint}$-$\mathtt{emb}$,
and $\mathtt{dir}$-$\mathtt{seq}$-$\mathtt{emb}$ approaches as
a function of the number of SFCs to embed.}
\end{figure}

Figures~\ref{fig:RES_Compare_Cost} and \ref{fig:RES_Compare_Time}
show respectively the cost and the required computing time for different
number of SFCs belonging to Slice of type~1 to be embedded (ranging
from $2$ to $10$). The embedding cost reflects the amount of infrastructure
node and link resources used for embedding these SFCs. The proposed
methods, \textit{i.e}., $\mathtt{prov}$-$\mathtt{joint}$-$\mathtt{emb}$
and $\mathtt{prov}$-$\mathtt{seq}$-$\mathtt{emb}$, , have similar
cost performance as that of the direct embedding, \textit{i.e}., $\mathtt{dir}$-$\mathtt{joint}$-$\mathtt{emb}$
and $\mathtt{dir}$-$\mathtt{seq}$-$\mathtt{emb}$. Nevertheless,
as depicted in Figure~\ref{fig:RES_Compare_Time}, the proposed approach
is faster than a direct embedding, when either performing in a joint
or sequential fashion. The difference increases with the number of
SFCs to embed. Note that in the proposed approach (\textit{i.e}.,
$\mathtt{prov}$-$\mathtt{joint}$-$\mathtt{emb}$ or $\mathtt{prov}$-$\mathtt{seq}$-$\mathtt{emb}$),
the computing time for the provisioning step has been taken into account.

\section{Conclusions\label{sec:Conclusions}}

This paper considers the problem of infrastructure resource provisioning
for network slicing in future mobile networks. Contrary to previous
best-effort approaches where SFCs are deployed sequentially in the
infrastructure network, here infrastructure resources are provisioned
to accommodate slice resource demands. For that purpose, a graph of
Slice Resource Demands is defined on the basis of the SLA between
an SP and the MNO. This graph describes the aggregated resource requirements
of the SFCs that will be deployed by the MNO for a given slice

Adopting the point of view of the InP, one tries to minimize the cost
related to the usage of the network infrastructure, in particular
the radio access network, while satisfying radio coverage constraints,
to ensure a minimum data rate for users in the geographical areas
where services have to be made available. This problem is cast in
the framework of MILP problem.

A two-step approach is proposed to address the complexity of this
problem. Radio resources on RRH are provisioned first to ensure the
satisfaction of the coverage constraints. Other constraints as defined
by the SRD graph are then considered. When resources have to be provisioned
for several concurrent slices, two variants have again been considered.
At each step, constraints related to each slice may be considered
either sequentially, or jointly. Due to the exponential worst-case
complexity in the number of variables of the MILP, as expected, sequential
methods are shown, through simulations, to better scale to network
topologies of realistic size. The price to be paid is a somewhat degraded
link utilization and a higher provisioning cost compared to the joint
approach. When both coverage and infrastructure network constraints
have to be taken into account simultaneously, \emph{i.e.}, the $\mathtt{JRN}$
approach, a minimum provisioning cost could be achieved, but this
approach requires a much larger time complexity than the four variants
of the suboptimal $\mathtt{CARP}$.

Once resources have been provisioned, the approach introduced in \cite{Riggio2016,Bouten2017}
may be used to deploy SFCs, but considering only a simplified infrastructure
network reduced to the nodes and links which have provisioned resources.
Simulations show that provisioning and then deploying is more efficient
in terms of computing time than direct SFC embedding.

Only static provisioning is considered in this paper. Resource provisioning
was done for a given time interval specified in the SLA over which
the service characteristics and constraints are assumed constant and
compliant with the variations of user demands within a slice. A level
of conservatism in the amount of provisioned resources is then required
to satisfy fast fluctuating user demands. One could imagine adaptive
SLAs to meet more closely the actual demands. The SLA may consider
several time intervals over each of which the service characteristics
and constraints are assumed constant, but may vary from one interval
to the next one. On the other hand, one could imagine that already
allocated SFCs may be updated during the lifetime of the slice. Adaptive
SLAs and dynamic provisioning techniques will be considered in future
work.

\vskip -2\baselineskip plus -1fil
\bibliographystyle{IEEEtran} 
\bibliography{ref_tnet}

\end{document}